\documentstyle[12pt]{article}

\newcommand{\rf}[1]{(\ref{#1})}

\renewcommand{\thefootnote}{\fnsymbol{footnote}}

\newcommand{\newsection}{    
\setcounter{equation}{0}
\section}

\def\appendix#1{
  \addtocounter{section}{1}
  \setcounter{equation}{0}
  \renewcommand{\thesection}{\Alph{section}}
  \section*{Appendix \thesection\protect\indent \parbox[t]{11.715cm} {#1} }
  \addcontentsline{toc}{section}{Appendix \thesection\ \ \ #1}
  }
\def \diag {{\rm diag}}
\def \Q{{\td Q}}
\def \HH {{\td H}}
\def \STr {{\rm STr }}
\def \Str {{\rm Str }}
\def \N {{\cal N}}
\def \SSTr{  {\widehat {\rm STr} } }

\newcommand{\tr}[1]{\:{\rm Tr}\,#1}

\def \Ho{H_{4({\bf 1})}}
\def \Ht{H_{4({\bf 2})}}
\def \Htr{H_{4({\bf 3})}}
\def \No{N_{4({\bf 1})}}
\def \Nt{N_{4({\bf 2})}}
\def \Ntr{N_{4({\bf 3})}}
\def \Noo{N_{0({\bf 1})}}
\def \Not{N_{0({\bf 2})}}
\def \Notr{N_{0({\bf 3})}}
\def \Vfo{V_{4({\bf 1})}}
\def \Vft{V_{4({\bf 2})}}
\def \Vftr{V_{4({\bf 3})}}
\def \Vo{V_{2({\bf 1})}}
\def \Vt{V_{2({\bf 2})}}
\def \Vtr{V_{2({\bf 3})}}
\def \Tr {{\rm Tr }}
\def \ha{{\textstyle{1\over 2}}}

\def \four{{\textstyle {1\ov 4}}}
 
\def\bi {\bibitem}
\def \ep{\epsilon}

\def \om {\omega}

\def \tF {\tilde \F}
\def \T {\tilde T}

\def \F {{\cal F}}
\def \g {\gamma}
\def \del {\partial}

\def \ha{{\textstyle{1\over 2}}}

\def \a {\alpha}
\def \b {\beta}
\def \chi {\chi}
\def \s {\sigma}

\def \m {\mu}

\def \td {\tilde }

\def \ci {\cite}

\def \C {{\cal C}}

\def \inv {^{-1}}
\def \ov {\over }
\def \four{{\textstyle{1\over 4}}}

\def \V {{\cal V}}

\def \pa {\Vert}
\def \hal{{{1\over 2}}}
\def \DD {{\cal D}}
 \def \G {{\Gamma}}

\def \foot{\footnote}

\def\np {{  Nucl. Phys. }}
\def \pl {{  Phys. Lett. }}
\def \mpl {{ Mod. Phys. Lett. }}

\def \pr  {{ Phys. Rev. }}

\def \cqg {{ Class. Quant. Grav. }}
\def \bi{\bibitem}

\def \T {{\td T}}

\def\det{\hbox{det}}
\def\be{\begin{equation}}
\def\ee{\end{equation}}
\def\beq{\begin{equation}}
\def\eeq{\end{equation}}
\def\bea{\begin{eqnarray}}
\def\eea{\end{eqnarray}}

\def \la{\label}
\def \ci{\cite}

\def \ov {\over}
\def \tr {{\rm tr}}
\def \Tr {{\rm Tr}}
\def \gym {g_{\rm YM}} 
\def \GG {{\bf \G}} 
\begin{document}
\begin{titlepage}
\begin{flushright}
ITEP--TH--40/97\\
ITP-SB-97-54\\
Imperial/TP/96-97/60\\
hep-th/9709087\\
\end{flushright}
\vspace{.5cm}

\begin{center}
{\LARGE  
 Long-distance interactions  of branes: 

%
correspondence  between  supergravity
 
and super Yang-Mills  descriptions  }\\
\vspace{1.1cm}
{\large I. Chepelev${}^{{\rm 1,}}$\footnote{
E-mail: chepelev@insti.physics.sunysb.edu }
and A.A. Tseytlin${}^{{\rm 2,}}$\footnote{Also at Lebedev Physics
Institute, Moscow. \ E-mail: tseytlin@ic.ac.uk} }\\
\vspace{18pt}
${}^{{\rm 1\ }}${\it ITEP,
Moscow, Russia and ITP, SUNY at Stony Brook, NY11794-3840}\\
${}^{{\rm 2\ }}${\it Blackett Laboratory, 
  Imperial College, London SW7 2BZ, U.K.}
\end{center}
\vskip 0.6 cm

\begin{abstract}
We address the issue of correspondence between classical 
supergravity and quantum super Yang-Mills (or Matrix theory) 
expressions for the long-distance, low-velocity  interaction potentials 
between  0-branes and  bound states of branes. The leading-order potentials 
are reproduced by the $F^4$ terms in the 1-loop SYM effective action.
Using self-consistency considerations, we determine a universal combination of 
$F^6$ terms in the 2-loop SYM effective action that corresponds to  
the subleading terms in the supergravity potentials in many cases,
including 0-brane scattering off 1/8 supersymmetric  $4\bot 1\pa 0$ and $4\bot4\bot4\pa0$ bound states representing extremal $D=5$ and $D=4$ black holes.  We give explicit descriptions of  these configurations in terms of 
1/4 supersymmetric SYM  backgrounds on  dual tori. Under a proper choice
of the  gauge field backgrounds, the 2-loop $F^6$  SYM action reproduces
the full expression for the subleading term in the  supergravity potentials, including  its subtle $v^2$ part.
\end{abstract}

\end{titlepage}
\setcounter{page}{1}
\renewcommand{\thefootnote}{\arabic{footnote}}
\setcounter{footnote}{0}


\newsection{Introduction}
One of the remarkable consequences 
of the open string theory   description  of D-branes  \ci{poll} 
is the existence of a close  correspondence  between the 
supergravity and  super Yang-Mills theory  results for 
certain interactions of D-branes and their bound states
\ci{poll,bachas,dkps,lifs1,bfss,liff1,ab,lifmat,lif3,ber,ballar,pp}.
For configurations of branes with  enough  amount of
underlying supersymmetry,  the 
long-distance and short-distance limits of the  string-theory 
potential  given by the annulus diagram are the same,  implying that  the 
 leading-order 
(long-distance) interaction potential  determined by 
the    classical 
supergravity limit of the closed string theory 
is  the same as the  (short-distance) one-loop potential 
produced by  the massless  open string theory modes, i.e. by  the  super Yang-Mills theory \ci{dkps}.
This was demonstrated  explicitly for the leading-order terms in the
 potentials
of interactions of 0-branes with  1/2 supersymmetric (non-marginal)
bound states \ci{ab,lifmat,lif3} and  with 
 1/4 supersymmetric marginal bound states \ci{CT1,CT2,GR}.
Similar conclusion was reached for the leading-order interaction of D-brane
probes with  1/8 supersymmetric bound states representing $D=5$ black holes \ci{dps,malda}.

On the SYM side, all of  the {\it leading-order}  
potentials (including also the  cases of interaction with
non-supersymmetric bound states of branes like $6+0$ \ci{lifs1,shein,
taylor, kraus,pie}, $8+0$ \ci{pierr} and  configurations corresponding to near-extremal $D=5$ black holes \ci{malda}) 
may be  obtained  by  plugging the corresponding  SYM backgrounds
into   the  leading universal  
$F^4$ terms in the IR part of the  one-loop  SYM
effective action in $D < 10$. 
The general  form of these 
$F^4$ terms was discussed in detail in \ci{MT,CT2,malda}. 

The  aim  of  this paper  will be 
 to attempt to understand if   the supergravity-SYM correspondence 
extends  also 
to  the level of  {\it subleading} terms in the interaction potentials.
The first  step  in this direction was made  in  \ci{BB,bbpt}
where  the   0-brane - 0-brane interaction 
was considered.
 It was shown that the  2-loop effective action in the 
$D=1+0$ dimensional reduction of 
SYM theory  computed for the relevant (velocity $v$, distance $r$)
background  has $v^6/r^{14}$ as the leading IR term (i.e.
 does not contain a  $v^4$-term \ci{BB}) and its coefficient 
is in precise agreement \ci{bbpt} 
with  the first subleading term in the supergravity potential
(computed using large $N$ limit or the `null reduction' prescription 
\ci{bbpt}  implementing  the suggestion of \ci{suss}).

We expect that as in the case of the leading  $v^4/r^7$ potential \ci{dkps}, 
this coincidence  should have  a weak-coupling 
 open string theory explanation 
and thus should be more universal, i.e. should  apply 
 also to  other appropriate  configurations of  branes
 of different  dimensions
and  with various amounts of supersymmetry
(in particular, to a  Dp-brane bound to other branes 
interacting with  a Dp-brane 
bound to  the same or different  combination of branes, 
and  to T-dual configurations). 
Moreover, similar relations  may   be true also between  all higher
order terms in the classical 
 supergravity potentials and the  leading infra-red  (low-energy) 
contributions
to the higher-loop  terms in the  SYM effective action in $D<10$ dimensions.

Like  the  $v^4/r^7$ potential originates from  the 
leading   IR term $F^4/M^7$ 
 in the 1-loop SYM effective action (with  an IR cutoff $M\sim r$), 
the  $v^6/r^{14}$ potential may be  related to  the  
leading  IR term $F^6/M^{14}$  in the 2-loop SYM effective action.
We  shall conjecture that in general\foot{While  our discussion  may   have 
 obvious  implications for the Matrix theory proposal \ci{bfss}, 
 we shall be assuming the weak string coupling limit
and consider only perturbative SYM contributions.}

(i)  the leading IR part (which   has  also an 
appropriate scaling with  $N$)  
of  the $L$-loop  term in the  $U(N)$ 
SYM effective action in $D=1+p$ dimensions  
has  a universal   $F^{2L+2}/M^{(7-p)L}$ structure;

(ii) computed for a SYM background  representing 
a configuration of interacting branes, 
 the  $F^{2L+2}/M^{(7-p)L}$
  term  should reproduce the  $1/r^{(7-p)L}$ term 
in the corresponding  long-distance classical supergravity  potential.

The first part of this  conjecture 
 is known to be true for $L=1$, and  we interpret the  
 result of \ci{BB} about the vanishing of the $v^4$
term in the 2-loop  $D=1$ SYM effective action as suggesting that
 it  is also true in general for $L=2$. The assumption that 
the leading part of the 2-loop term is the $F^6$  one  is also 
implied  by the $F^4$ term  non-renormalisation theorem of 
\ci{seid}.\foot{Though  `universality' or `BPS saturation' of terms  
 with 
higher than 6 powers of $F$ may seem less plausible, there are, in fact, 
string-theory examples  of  higher-order terms that receive contributions
only from one particular  loop order, to all orders 
 in loop expansion \ci{ant}.}
While   we formulated the above  conjecture for general $L$,  most of 
considerations in this paper  will be restricted to the $L=2$ case.

Given that direct computation of the $L  >  1$ terms in the  $D >1$ 
SYM effective action is hard, 
  our strategy in trying to test this  conjecture will be 
 to make  a  plausible assumption
about the structure  of  the relevant part of  the SYM effective action
and then  check  if our ansatz 
 can match  known  supergravity  potentials 
for various special configurations of branes. Since  different 
brane systems with different amounts of supersymmetry 
have very different SYM descriptions, the  conjecture that 
all of the corresponding interaction potentials originate from 
 a single universal  $F^{2L+2}$-type SYM  expression  
provides
highly non-trivial constraints on the latter.

We shall  study  the first non-trivial  case of $L=2$ 
and demonstrate that indeed the interaction potentials  for  
various examples of interactions of 0-branes with  type IIA BPS 
bound state of branes with $1/2,1/4$ or $1/8$  of supersymmetry
and non-trivial 0-brane content 
can be described by a certain  universal combination of  
$F^6$ terms 
in the 
leading IR part of the  2-loop SYM effective action in $D=1+p$, 
thus suggesting a   non-trivial generalisation of the
 0-brane -- 0-brane result of 
\ci{bbpt}. 

The type IIA 
brane systems  we shall  deal with   will be  of the following  special 
 type:
a 0-brane probe (a cluster of $n_0$ 0-branes) interacting with a  BPS 
(marginal or non-marginal, $1/2^n$ supersymmetric) 
bound state of branes  having 
 a non-zero 0-brane charge ($N_0$) component 
and  wrapped over a torus $T^p$. 
The probe will have  a velocity  along 
a direction transversal  to the `internal' torus. 
T-duality  along all of the directions of $T^p$ relates this  system 
to a system of 
Dp-brane probe (with charge $n_0$)
 parallel to  a bound state of $N_0$ Dp-branes 
bound to Dq-branes ($q<p$) and  wrapped over the dual torus $\td T^p$.
Such system should  thus 
have a $U(n_0 + N_0)$ SYM   description  \ci{wit2}
with a non-trivial SYM 
background reflecting 
 the presence of other branes in the bound state \ci{doug} and the 
 velocity of the probe \ci{bachas,cakl}. 

On the  supergravity side, the  action  for a  0-brane probe  interacting 
with a background produced by a marginal bound state of branes ($1\pa 0, \ 4\pa 0$, \ 
$4\bot 1 \pa 0$ or $4\bot 4\bot 4 \pa 0$)  which is essentially the same as  the  action 
for a $D=11$ graviton scattering off an M-brane configuration
($2+$wave, $5+$wave, $2\bot 5+$wave or $5\bot5\bot5+$wave)
has the following general structure
\ci{ts3,ts4,dps}
\be 
I_0 = -T_0 \int dt\ H_0\inv \bigg[\sqrt { 1 - H_0 H_1...H_k v^2} -1 \bigg]
\equiv \int dt \ \bigg[ \ha T_0 v^2 - \V (v,r) \bigg] \ .
\la{pou}
\ee
Here $H_0$ and $H_1, ..., H_k $ ($k=1$ for 1/4 supersymmetric bound states and
$k=2$ or $k=3$ for  1/8 supersymmetric bound states)  
   are the  harmonic functions  $H_i = 1 + Q_i/r^{7-p}$
corresponding to the constituent charges  of the bound state.
Since  for D-branes  $Q_i \sim g_s N_i$ and $T_0 \sim  n_0 g_s\inv$ where 
$g_s$ is the string coupling constant,  the 
long-distance expansion of the classical 
supergravity interaction potential $\V$  has the following form\foot{For simplicity, we are assuming here that the  bound state has only RR charges;
cases with non-vanishing  fundamental string charge or momentum
  ($\Q_1 \sim g^2_s$) 
 can be treated in a similar way, see  \ci{dps,GR} and  section 3.1 below.}
\be
 \V  = \sum_{L=1}^\infty \V^{(L)} = 
{ n_0 \ov g_s}  \sum_{L=1}^\infty 
 \bigg({ g_s  \ov r^{7-p}}\bigg)^{L}  {k_L} (v, N_i)  
\ ,  \la{vvv}
\ee
so that  the $(1/r^{7-p})^L$ term 
has the same $g_s$ dependence as in the $L$-loop
term in SYM theory with coupling $\gym^2 \sim g_s$.

The detailed form of the coefficients $k_L$ in the
 potential in \rf{vvv}  reflects the two important features
of the 
 action \rf{pou}: \ 
 (1) the  special role played by the 0-brane function $H_0$, and \ \ (2)
the appearance of the product of the `constituent' harmonic functions.
The second property  is  the  direct consequence of 
 the `harmonic function rule' structure \ci{ts2}
of the  supergravity backgrounds representing marginal 
BPS bound states of branes. It implies that all the 
 constituent charges,  except the 0-brane one, 
  enter $\V$ in a  completely symmetric way. 

The  asymmetry between $H_0$ and $H_1,...,H_k$  is strengthened
by  further important assumption
that in expanding  \rf{pou} in powers of $1/r^{7-p}$ one should
take  $H_0$  without the usual asymptotic term 1, i.e.  as 
$
H_0 =  Q_0/r^{7-p}, \ Q_0 \sim g_s N_0$. 
This prescription  which is crucial for the 
precise correspondence with the SYM theory 
already at the leading (1-loop)  level  
may be interpreted  at least in  two possible  ways.
One may  assume (as  was done in \ci{lifmat,lif3,CT1}) 
that $N_0$ is large  for fixed $r$ and $g_s$ 
(in particular,  $N_0 \gg N_1,..., N_k$),
 so that $H_0= 1 + Q_0/r^{7-p} \approx Q_0/r^{7-p}$. 
Alternatively, one may keep $N_0$ finite but consider  the $D=10$ 
brane system  as  resulting  from an M-theory configuration 
 with 
 $x^- = x^{11} -t$ direction being compact \ci{suss}.  As  was 
pointed out  in \ci{bbpt}, the 
dimensional reduction of $D=11$ gravitational
wave combined with  M-brane configurations
along $x^-$  results in supergravity backgrounds with 
$H_0 \to H_0 -1= Q_0/r^{7-p}$.  In what follows we shall always  set
$H_0 =  Q_0/r^{7-p}$ without making any  assumptions about magnitude of 
 $N_0$.  

To reproduce  the detailed form of the subleading terms in 
\rf{vvv} from  the   SYM theory 
it should be possible to encode the  structure of the 
supergravity  expression \rf{pou} 
 (in particular, the cross-terms coming from the product $H_1 ...H_k$
of the harmonic functions and reflecting effective 
 interactions of constituent branes in the bound state) in the 
explicit 
form of \  (a)  universal  $F^{2L+2}$ terms in the 
 SYM effective action, \ 
  and \  (b)  specific SYM background  representing the  
bound state of branes   on the SYM side.

A  $U(N_0)$ SYM background $F_{ab}$  on the dual torus $\td T^p$ 
which is a candidate for a description  of   a marginal BPS 
 D-brane bound state
with $1/2^n$ of  $\N=2,D=10$ supersymmetry  should satisfy the 
following conditions:   
(1) $F_{ab}$ 
 should  
 preserve  $1/2^{n-1}$  of ${\cal N}=1, D=10$ supersymmetry of SYM theory;\ \
(2) considered as a gauge field background on a Dp-brane world volume, 
$F_{ab}$ should induce \ci{doug}
 only the  charges $\sim\int \tr ( F\wedge...\wedge F$)
of  constituent branes; \ \ 
(3) the classical D-brane  (Born-Infeld)
 action computed  in  this background 
\be
I = T_p  \int d^p \td x \ \Str \sqrt {- \det ( \eta_{ab} I +  F_{ab} ) }  \ , 
 \la{born}
\ee
should reproduce the mass of the corresponding 
marginal BPS bound state
 which is in agreement with the supergravity background describing the bound state, 
i.e. is  proportional to the sum of  
charges of the constituent branes. 
In general, these conditions do not  fix the required  SYM background uniquely. 
One of the  lessons  of  our  discussion below 
will be the crucial role of  an appropriate 
 choice of the   SYM representation of the brane bound states
  for
supergravity -- SYM  correspondence at the subleading level.

To fix the form of the relevant  2-loop $F^6$  term in the SYM effective
action $\G$  we shall proceed in  steps,  first 
considering the interaction of a 0-brane probe 
 with  1/2 supersymmetric non-marginal bound state ($p+...+0$)
 and then 
turning to more complex cases of interaction with 1/4 and 1/8 
supersymmetric bound states.
Demanding the  agreement  with the  supergravity potential $\V$ 
 in the simplest cases
we will be able  to extract the
 information about the required  structure of $\G$ 
which  will then be  checked and sharpened (by  including 
 terms that 
were vanishing on  the previous less complicated  
 backgrounds) 
 on more complex  examples of  0-brane scattering off 
 bound states with  reduced   amount of supersymmetry.
This procedure will  turn out to be by 
 far less  arbitrary as it may   seem first
 and  the emerging consistent picture will provide  support
  for 
the above  conjectures.


In section 2 we shall  describe  an  ansatz for  
 the leading IR part $\G$ of the full SYM effective action $\GG$ 
which is expected to  reproduce the supergravity potential  $\V$ \rf{vvv}.
The  proposal for  the  higher-loop 
terms in  $\G$, and, in particular, for the  2-loop  $F^6$ term, 
which we shall make  will be  motivated   and tested
by comparing with  subleading terms in  $\V$  
in various special cases
in  the following sections.

In section 3 we shall consider the interaction  between 
a 0-brane  and  a non-marginal 1/2 supersymmetric
bound state ($p+ ...+0$) of type IIA 
D-branes. Since the action for the latter treated as a probe 
can be described by switching on a constant abelian background
on the Dp-brane world sheet, it will be straightforward to 
demonstrate that the  corresponding 
supergravity   potential $\V$   admits a 
 SYM interpretation, suggesting as a result the required  pattern 
of  the Lorentz-index  (BI polynomials)  and internal index (single trace
in adjoint representation $\Tr$) contractions in $\G$.  As a particular example, 
we shall consider 
the 0-brane - 0-brane  scattering,
 interpreting
the 2-loop result of \ci{bbpt} as a special case of the general $\Tr F^6$ SYM
expression and suggesting its extension to all loop orders.


 The 0-brane interactions with 1/4 supersymmetric marginal bound states 
(of a  fundamental string and a 0-brane $1\pa 0$ and 
of a 4-brane and a 0-brane  $4\pa 0$) 
  will be discussed in section 4.
We shall find that exact all-order 
 expression for the classical  $0- (1\pa 0)$ supergravity potential
is reproduced by the same ansatz for the SYM effective action 
that was  giving  the full $0-(p+...+0)$ potential in section 3.
The situation in the $0-(4\pa 0)$  case turns out to be  more subtle as 
the corresponding gauge field background is 
described by two  different (though still commuting)  $su(N)$ matrices. 
We shall determine  an extra  correction term 
in the 2-loop part of  $\G$ which  was vanishing in the previous
$0-(p+...+0)$ and $0-(1\pa 0)$ cases 
but is necessary for the  exact  agreement 
between the subleading term $\V^{(2)}$ in \rf{vvv}
and the $O(F^6)$ 2-loop SYM  term  
in the $0-(4\pa 0)$ case.

The consistency of the emerging  expression 
 for the 2-loop SYM effective action
will be  tested further 
in section 5 where we shall  consider the 0-brane interactions
with 1/8 supersymmetric  marginal bound states $4\bot 1\pa 0$ (or $5\bot 2+$wave in $D=11$ ) and 
 $4\bot 4\bot 4 \pa 0$  (or $5\bot 5\bot 5+$wave in $D=11$),  
 which (when wrapped over $T^5$ and 
 $T^6$)
 correspond
to $D=5$  and  $D=4$ extremal black holes with regular horizons.
 These  bound state configurations  are represented by curved type IIA  $D=10$ 
supergravity backgrounds but also 
admit  simple descriptions in terms of 1/4 supersymmetric  SYM 
gauge field backgrounds on the dual tori, which, as we shall see, 
 are not unique.  We shall find that in the $4\bot 1\pa 0$ ($D=5$ black hole) case
there exists a natural choice 
of a SYM background on $\td T^5$ which when 
 substituted into the 2-loop SYM 
action   determined in the previous sections reproduces the complete 
expression for the subleading term $\V^{(2)}$  in the supergravity potential, 
including the $v^2$ term (cf. \ci{dps}). 
The   $4\bot 4\bot 4 \pa 0$   ($D=4$ black hole)  configuration 
will  be represented  either  by  commuting  ($[F, F]=0$) 
or non-commuting (`three instanton')  1/4 supersymmetric
gauge field backgrounds on $\td T^6$. 
We shall find that to reproduce the full expression for the 
 subleading  supergravity potential $\V^{(2)}$
it is necessary to use the non-commuting  SYM background and to include
in the 2-loop  effective action   
 terms with commutators of $F$. Such commutator terms  should, in general, 
 be present in $\G$ but were  not contributing 
 in  our  previous examples  which were 
described by commuting SYM backgrounds.

Some important  remaining  questions  will be  mentioned in section 6. 
In Appendix  we shall describe another  `commuting' 
representation for the $D=4$ black hole configuration.


\newsection{Interactions between branes and 
Super Yang-Mills   effective action}
Our aim  in this section will be to describe   possible 
structure of the leading IR part $\G$  of the full SYM effective action $\GG$
which is  relevant for the discussion of 
interaction potentials between a 0-brane probe and a 
bound state of branes
wrapped over a p-torus.  
We shall  consider the case of weak string coupling   and 
assume that   such  configurations  can be represented  by  appropriate
backgrounds  in SYM theory in $D=p+1$ dimensions.
The  proposal for  the  higher-loop 
terms in  $\G$, and, in particular, for the  2-loop  $F^6$ term, 
we shall make below will be  motivated   and tested
by comparing with the  subleading terms in the 
supergravity interaction potentials in various special cases considered 
in  the following sections.
Our  ansatz for $\G$ 
will be   a natural generalisation of the one-loop $F^4$-term in $\GG$
 which governs the leading-order interaction potentials between different
combinations of branes \ci{CT2}
and   in the special 0-brane scattering  case  it will   also agree 
 with  the  direct 2-loop $D=1$ SYM  calculations in 
\ci{BB, bbpt, GGR}.

 The fields of the maximally supersymmetric  $D$-dimensional SYM 
 theory  (obtained by  reduction from $D=10$ SYM) 
are  the vectors $A_a$ ($a=0, ..., D-1$) and
the scalars  $X_i$ ($i= D, ...,9$).   
 In general,  both may have  non-trivial  background values.
The system we  will be interested in  consists of 
a $0$-brane  probe (a marginal bound state of $n_0$ 0-branes)  
interacting with a BPS bound state of branes 
containing, in particular,  $N_0$\  0-branes,  and 
 wrapped over $T^p$.
  Under  T-duality  along all of the 
directions of the torus 
it becomes a Dp-brane probe with charge $n_0$ 
parallel to a Dp-brane  source with charge $N_0$ bound to some
 other branes of lower dimensions. 
Assuming that the probe and the  source are separated by a distance $r$ 
in  the direction 8 and that probe has velocity $v$ along the transverse 
 direction 9, 
this configuration  may 
   be described by the following  $u(N)$, $N=n_0 + N_0$, SYM background
on the dual torus $\td T^p$  
\be
\hat {A}_a=
 \left(
 \begin{array}{cc}0_{n_0\times n_0}  & 0 \\ 0 & {A}_a \\ \end{array}
 \right) , \ \ \ \ \hat {X}_i= \left( \begin{array}{cc}
0_{n_0\times n_0}  &0 \\ 0 & {X}_i \\ \end{array}  \right) , \ \ \ \  i\not=8,9 \ ,   \la{baccc}\ee
\be
\hat X_8=
 \left(
 \begin{array}{cc} r\ I_{n_0\times n_0}   &  0 \\ 0  &  0_{N_0\times N_0} \\
 \end{array}
 \right), \ \ \ \ 
 \hat X_9= \left( \begin{array}{cc}  vt\ I_{n_0\times n_0}   & 0  \\   0 & 0_{N_0\times N_0} \\ \end{array} \right) ,
\la{becc}
\ee
 where ${A}_a$ and $X_i$ are $N_0 \times N_0$ matrices in the 
fundamental representation of $  u(N_0)$
 which  describe the source bound state.
For example, $A_m$ may be an instanton field representing  
 a 0-brane charge on a  4-brane (i.e. $4\pa 0$ bound state) \ci{doug,vafa}
while $X_i$ may be a wave  representing a momentum flow
 along some direction of $\T^p$, or,   
after T-duality along that direction,  a fundamental string charge
(for $p=1$ this corresponds to  the $1 \pa 0$ bound state) 
\ci{imam,wu,gopak}.

In general, the SYM effective action  $\GG$
(computed  using the  background field gauge)
is a gauge-invariant  functional
of the background  fields  $\GG(A,X)= \GG (F,X,DF, DX, ...)$. 
Assuming that  the  source brane 
configuration is a supersymmetric (BPS) one, $\G$ should 
vanish for $v=0$ as well as for $r\to \infty$.
  The long-distance interaction potential  will be 
 given  by the  low-energy 
expansion of $\GG$ in powers of  $F$ multiplied by powers of 
$1/X_8$ or $1/r$.
The background value of $X_8$  plays the role of an effective 
IR cutoff
(in the open string theory picture it is related 
to  the mass of  the  open string states stretched between the probe and source branes).\foot{Though not all of the 
SYM excitations are getting  explicit mass terms, the remaining IR divergences must
cancel out  as in  \ci{BB}  and  should  not contribute to the 
leading IR  (`interaction potential' $\G$)
part  of $\GG$.}

The  dependence on  derivatives of the scalar fields 
$X_i$ (and thus, in particular, on $v$  \ci{bachas,ber,CT2})
may be formally determined from the dependence of $\GG$ on the gauge field
 in a  higher-dimensional background  representing 
T-dual ($X_i \to A_i$)  configuration.
Indeed,  from the point of view of the open string theory description of 
D-branes \ci{poll}  $\GG$ 
 should   be  related to  
the  short-distance limit of the open string  
loop diagrams and thus should be 
`covariant' under the  T-duality \ci{poll, bachas}
(the string partition function is given by the path integral with the source term 
$\int dt [ \del_t x^a A_a (x) + \del_n x^i X_i (x) ]$
so that  $A_s \leftrightarrow   X_s$ under T-duality).  
The dependence on $\del_m X_i$ 
 may thus be  determined  from  
$\GG (F)$ in a higher-dimensional pure gauge field 
 background  with $F_{mi} = D_m X_i$.

The problem is  then formally 
reduced to 
finding the  SYM effective action 
 in the case of  a purely  gauge field 
background
 and  with an effective 
 IR cutoff $M=r$  provided by the scalar field background 
(we  
set the string tension $T= (2\pi \a')\inv=1$).\foot{We shall be interested only in the 
low-energy  limit of the SYM  theory, i.e. will not  consider the 
 UV  cutoff dependent parts in the corresponding effective 
actions (assuming  the existence of 
an explicit UV cutoff effectively provided in  the weak-coupling  case
by the string theory).}
The corresponding 
 SYM theory on $\T^p$ has the  action  
\be
 S = - { 1 \ov 4 \gym^2}  \int d^{p+1 } \td x\  \tr F^2 + ... 
= - { 1 \ov 8  \gym^2  N} \int d^{p+1 }  \td x\  \Tr F^2 + ...\  , \ \ \ \ 
d^{p+1} \td x \equiv dt d^{p} \td x \ , 
 \la{yyxx}
\ee
\be \gym^2 = (2\pi)^{-1/2} g_s \td V_p \   , \la{coe} \ee
where $g_s$ is the string coupling constant,
$\td V_p = \int d^p \td x $ is the volume of $\T^p$ 
($V_p \td V_p = (2\pi/T)^p$) 
  and $\tr$ and $\Tr$
are the traces in the fundamental and adjoint representations of $su(N)$.
The value of the SYM coupling  constant  
is dictated by  T-duality considerations
for  a system of 0-branes on a torus
 \ci{bfss,tay,grt,sussk}. 
 The  action for a collection of  Dp-branes wrapped over $\td T^p$ is  
$S= - T_p  \int d^{p+1 }  \td x\ \ \tr \sqrt{-\det(\eta_{ab} + \del_a X^i \del_b X^i 
+ T^{-1}  F_{ab})} + ...$,  where $T_p$ is the Dp-brane tension \ci{poll}. 
Viewed as the leading term  in 
 this action, eq. \rf{yyxx}  should  thus have  the  coefficient  
$ \gym^2 = T^2/T_p =  (2\pi)^{(p-1)/2}  T^{(3-p)/2} \td g_s$, where $\td g_s$
is the string coupling constant of the T-dual theory  satisfying  the standard relation 
 $ V_p/ g^2_s = \td V_p /\td g_s^2$. For $T=1$ this gives \rf{coe}.

In $D=p+1$ dimensions $\gym^2$ has mass dimension $3-p$ so that
on dimensional grounds, 
the relevant part of $\GG$ expanded in powers of  $ F$ should 
have the following structure 
\be
\GG =   \sum_{L=1}^\infty  (\gym^2 N)^{L-1} 
\int  d^{p+1} \td x  \sum_n  c_{nL} { 1 \ov  M^{2n - (p-3) L -  4}}  F^{n} 
\ , 
\la{stru}
\ee
where $L$ is the number of loops and 
$F^n$ stand for all possible contractions of $n$ factors
of the field strength matrix (we shall not explicitly  consider
 terms with covariant derivatives of $F$
assuming  that they  can be ignored  for the relevant backgrounds).

In what follows we shall be interested   only in  a  special   subset 
of terms in \rf{stru} 
(generalising the `diagonal terms' in \ci{bbpt}) 
which have the right coupling $g_s$, 0-brane charge $N_0$ 
 and distance $r=M$ dependence to be in correspondence with 
the terms in the long-distance expansion \rf{vvv}  of the  {\it classical} 
supergravity interaction potential $\V$ between  a Dp-brane probe
(with tension $\sim n_0/g_s$) and a Dp-brane source (with `charge' parameter
$\sim g_s N_0$).
As it is clear from the string-theory description 
 of  interaction between two Dp-branes, these terms should come out of {\it planar}  diagrams of SYM theory.
As  was already  mentioned  in the Introduction,  our  conjecture   is that 
due to  maximal underlying  supersymmetry of the SYM  theory, 
the terms $F^{2L+2}/M^{(7-p)L}$ represent, in fact,  the {\it leading} 
IR  contribution
 to  $\GG$ at   $L$-th loop order.
 This is true for $L=1$ \ci{fraa,MT}
and, in view of the results of 
\ci{BB,seid},   should certainly be the case  also for $L=2$.

The sum of these leading IR   terms   at each loop level 
 will be denoted by $\G$
\be
\G   =\sum_{L=1}^\infty \G^{(L)} =  \ha \sum_{L=1}^\infty  \int d^{p+1} \td x 
  \  \bigg( {a_p  \ov  M^{7-p}}\bigg)^L  
  (\gym^2 N)^{L-1} \  \hat  C_{2L+2} (F) \ , 
\la{eff}
\ee
$$ 
\hat  C_{2L+2} (F) \sim  { F^{2L+2} }   \ . 
$$
The coefficients $a_p$ here must be universal, i.e. they 
cannot depend on $N$.  As we shall find
 from comparison with the supergravity potential ($2\pi \a'=1$) 
\be
 a_p = 2^{2-p}\pi^{-(p+1)/2}
{\Gamma ({\textstyle{ 7-p \ov 2}})} \ .   
\la{coee}
\ee
The structure of the coefficients 
 $\hat  C_{2L+2}(F) $ should be such  that when  computed for the 
 relevant  gauge field backgrounds
 they  should 
contain an extra factor of $n_0 N_0$  so that the  order 
$n_0 N_0^L$ term  in \rf{eff} could match  the corresponding term in  the supergravity expression 
$ \int dt\ \V$   \rf{vvv}.

The central  question which we shall  be addressing  below 
  is the following: 
which    Lorentz 
and internal index structure of $\hat C_{2L+2}$ or the  `diagonal' $F^{2L+2}$ terms  in \rf{eff}
is required in order for the resulting $\G$ 
 to agree with the supergravity potential 
\rf{vvv}. Since there are several inequivalent 
 configurations  of branes  (involving   BPS bound states of branes 
with different amounts of supersymmetry),  the 
assumption  that the  interaction potentials
for all of them should be described by the same  universal 
SYM expression  $\G$ \rf{eff} imposes  
non-trivial  constraints
on the latter.

Let us first  assume that the background field  strength $F_{ab}$
belongs to the Cartan part
 of $su(N)$, i.e.  that all of its components commute, 
 $[F_{ab}, F_{cd}]=0$. This 
 will be the case for most of the examples discussed below.
Our main assumption  (motivated  by 
the  form of the supergravity potential 
in the case of   0-brane 
 interaction  with 1/2 supersymmetric non-marginal bound states
of D-branes discussed in section 3)
will be that  at least for commuting 
backgrounds, i.e. up to the `commutator terms' involving $[F,F]$,
  the structure of 
the Lorentz-index  contraction  in $\hat C_{2n} \sim F^{2n}$
in \rf{eff} is the same as in the  polynomials
 $C_{2n}=O(F^{2n})$ appearing in the  expansion of the  
abelian Born-Infeld action, 
\be
\sqrt{-\det{(\eta_{ab} +   F_{ab} })} 
= \sum_{n=0}^\infty     C_{2n} (F)   \ ,  
\la{see}
\ee
\be
C_0=1\ , \ \  \ \ \ C_2 = -  { 1\ov 4} F^2  \ , \ \ \ 
C_4 = - { 1\ov 8}
 \bigg[F^4 - { 1\ov 4} (F^2)^2 \bigg]  \ , 
\la{fff}
\ee
\be
C_6 = - { 1\ov 12} \bigg[F^6 - { 3 \ov 8 } F^4 F^2  
+ { 1 \ov 32} ( F^2)^3\bigg] \ ,  \ \ ....  \ ,  
\la{sixx}
\ee
where $F^k$ is the trace of the matrix product in Lorentz indices, i.e. 
$$  F^2 = F_{ab} F_{ba}\ , \  \ \  \  F^k = F_{a_1a_2}
 F_{a_2a_3} ... F_{a_ka_1} \ . $$
This is indeed what happens (for generic $F_{ab}$) 
  in  the explicitly 
 known  1-loop 
expression for $\G$ \ci{MT,CT2}
\be
\GG^{(1)} =-\frac{{\Gamma ({\textstyle{ 7-p \ov 2}})}
  }{ 2 (4\pi)^{(p+1)/2} M^{7-p} }\int d^{p+1} \td x \  {\bf b}_8 +O\big(\frac{1}{M^{9-p}}\big) = \G^{(1)} +  O\big(\frac{1}{M^{9-p}}\big) 
\ ,    \la{hh}
\ee
i.e. (cf. \rf{eff}) 
\be
\G^{(1)}
=   \frac{ a_p } {   2 M^{7-p} }\int d^{p+1} \td x \ \hat  C_4 (F)
\ , \    \la{hhh}
\ee
$$
\hat  C_4 = \STr\  C_4 = - {\textstyle{ 1 \ov 8}} {\bf b}_8
= - {\textstyle{ 1 \ov 8}} \STr \bigg[ F^4 -\four (F^2)^2 \bigg] 
$$
\be 
 = \ -  {{\textstyle{1\over 12}}} \Tr \bigg(F_{ab} F_{bc} F_{ad} F_{dc}
+ \ha  F_{ab} F_{bc}   F_{dc} F_{ad} - \four
F_{ab} F_{ab} F_{cd} F_{cd}  - {{\textstyle{1\over 8}}}
F_{ab} F_{cd}  F_{ab} F_{cd} \bigg) \,. 
\la{bbb}
\ee
Here $\STr$ is the symmetrised trace in the adjoint representation
of $su(N)$ which may be expressed 
 in terms of traces  $\tr$ in the fundamental 
representation  \ci{CT2}.

In general,  only the traceless $su(N)$ part of the $u(N)$
background field matrix $F$ couples to the quantum fields and thus enters 
the effective action.
Symbolically, if  $F= \pmatrix{ F_1 & 0 \cr 0 & F_2 \cr}, $ where
$F_i$ belong to $ su(N_i)$, $N=N_1 + N_2$, 
then $\Tr F^4 = \Tr_1 F^4_1 +   \Tr_2  F^4_2 + f(F_1,F_2)$, \ $f(F_1,F_2)=
2[ N_1 \tr_2 F_2^4 +  N_2 \tr_1 F_1^4 +  6 (\tr_1 F_1^2)  (\tr_2 F_2^2) ]
$  (see eq. (2.16) below) and it is the latter part $f(F_1,F_2)$ 
that describes
interaction (see also \ci{malda})\foot{We are grateful 
to J. Maldacena for a discussion of this point.}
 between two clusters of branes.\foot{In general,
 if $\hat F, \ \hat F_1$ and $\hat F_2$ are the traceless 
parts of $F, \ F_1$ and $F_2$ then 
the interaction potential is given by
$\Tr \hat F^4 - \Tr_1 \hat F^4_1 -   \Tr_2  \hat F^4_2$.}
The `self-energy' terms $\Tr_1 F^4_1$ and $   \Tr_2  F^4_2$ 
vanish in the case when $F_1$ and $F_2$ are  supersymmetric 
SYM backgrounds  representing BPS states of branes. 
Since this  is the case we will be discussing below, 
 we will not discriminate between $\G$ and interaction potential
 between branes.

Our next   assumption will be that 
for  {\it commuting } $F_{ab}$ backgrounds, 
the  pattern of  contraction over the internal indices in $\hat C_{2n}(F)$
should be  similar to that in  \rf{bbb}, i.e. to the 
 {\it  single}  (symmetrised) trace in the  {\it adjoint}  representation.
We shall  see that   the  simple ansatz 
\be
 \hat  C_{2n} (F) =  
 \STr \ C_{2n} (F) \ ,  
\la{eee}
\ee
where $\STr$ is applied to the polynomial 
$C_{2n}(F)$ appearing in the expansion of the  BI action \rf{see} with
each $F_{ab}$  now  promoted to an $su(N)$ matrix,\foot{For commuting $F_{ab}$ the symmetrised trace $\STr$  is 
  of course equal simply to  the trace   $\Tr$. 
The use of  symmetrisation here is  to isolate 
the terms that are non-vanishing on commuting $F_{ab}$ from the remaining `commutator terms' (cf. \ci{dbi}). The symmetrisation is also helpful
in order   to  express $\Tr$  in terms of 
traces in  the fundamental representation ($\Tr $ of a product of several different matrices  takes  simpler form   if one symmetrises 
the  factors  in the product).}
leads to  $\G$  \rf{eff}   which  reproduces  the {\it full} 
supergravity potential $\V$  in the case of  a 
$0$-brane interacting  with a general 
 1/2 
BPS  bound state $p+...+0$ (e.g., with another $0$-brane or $2+0$ bound state)
as well as in the case of  a $0$-brane interacting  with the 
1/4 supersymmetric bound state $1\pa 0$. The $\Tr$ structure  of $\hat C_{2n} $ 
 provides, in particular, the required $n_0 N_0$ factor. 
 
We shall find, however, that $\STr$ in  \rf{eee} 
should be modified by 
certain correction terms (which vanish in the above  special cases
where the background $F_{ab}$ is essentially
proportional to  a single $su(N)$ matrix)
but  whose presence   is  needed  for 
 the  SYM -- supergravity correspondence in the case
of more complicated backgrounds involving self-dual gauge field 
strengths  ($4\pa 0$ and its generalisations). 

The difference between the 1-loop and higher loop terms 
in $\G$  in what concerns the  internal index structure  
is clear from the form of the   
$SU(N)$ YM  Lagrangian  expanded ($A^s_a \to A^s_a + B^s_a$)
near a background $A^s_a$. In the 
  background  field gauge ($s=1,...,N^2-1$) 
$$ L =  {1 \ov g^2_{\rm YM}N } 
 \bigg[ \ha B^s_a \Delta^{sr}_{ab}  B^r_b + 
 f_{srt} B^s_a B^r_b  \DD_a B^t_b 
+ f_{srt} B^s_a B^r_b f_{s'r't} B^{s'}_a B^{r'}_b\bigg] \  ,    $$
where  $\Delta_{ab} = - \eta_{ab} \DD^2 -2 F_{ab} ,\  \DD= \DD(A)$.
The 1-loop effective action (obtained after $\Delta_{ab} \to 
\Delta_{ab} + M^2$) has, indeed,  the form of a  sum of 
$ 1/M^n$ terms  multiplied by a 
{\it single} trace in the  {\it adjoint} representation 
 of a  polynomial in  $F_{ab}$.
 This is no longer so in general for higher-loop corrections
to $\GG$  (there are many different contractions 
from products of the structure constants $f_{rst}$).
It  appears, however, that the contributions to 
the `diagonal' or leading IR  part $\G$  \rf{eff}  of $\GG$ \rf{stru}
do have a  $\Tr$-type
   structure, 
 up  to the `commutator terms' (i.e. up to the 
 terms that vanish 
  when evaluated on simple `commuting' 
backgrounds).  One may attempt to understand 
 this using large $N$  limit considerations.

While  the 1-loop coefficient  $\hat C_4$ \rf{bbb} is equal to 
$\STr C_4$  for generic $F_{ab}$,  
 we shall use the following ansatz  for  $\hat C_{2n}$ with 
higher  $n >2$  
\be
 \hat  C_{2n} (F) =  
 \SSTr \ C_{2n} (F)  = \STr \ C_{2n} (F)  +  \STr'C_{2n}(F) \  . 
\la{eeer}
\ee
Here  $\SSTr$   will  be defined  in terms of   somewhat
different  as compared 
to $\STr$   combination of  
symmetrised  traces in  the fundamental representation (see below). 
We shall explicitly determine $\SSTr$  or the `correction term' 
$\STr' C_{2n}$ in \rf{eeer} from comparison with supergravity  potential 
  for $n=3$, i.e. in the case of 
 the  2-loop coefficient
 $\hat C_6$.

For a single $su(N)$ matrix  $Y$
the traces in  the adjoint and fundamental representations 
are related as follows (see, e.g., \ci{patera})
\be
\Tr Y^2 = 2 N \tr Y^2\ , \ \  \ 
\ \  \Tr Y^4 = 2N \tr Y^4 + 6 \tr Y^2 \tr Y^2 \ ,  
\la{traac}
\ee
\be
 \Tr Y^6 = 2N \tr Y^6 + 30 \tr Y^4 \tr Y^2 - 20 \tr Y^3 \tr Y^3 \ . 
\la{trac}
\ee
Similar relations apply  to symmetrised products of $su(N)$ generators $Y_s$, 
e.g., 
$$
\STr (Y_{s_1}  ... Y_{s_6} )  = \Tr [Y_{(s_1}  ... Y_{s_6)} ] 
$$
\be
= 2N \tr [Y_{(s_1}  ... Y_{s_6)} ]
 + 30 \tr [Y_{(s_1}  ... Y_{s_4}] \tr [Y_{s_5} Y_{s_6)}]
 - 20\tr [Y_{(s_1}  ... Y_{s_3}] \tr [ Y_{s_4}... Y_{s_6)}] \ . 
\la{rtr}
\ee
We assume that $\hat C_6$ in \rf{eff}
is given by 
\be
\hat C_6 (F)  = \SSTr (Y_{s_1}  ... Y_{s_6} ) \ C^{s_1...s_6} (F) \ , 
\la{oonn}
\ee
\be
C^{s_1...s_6} (F)  \equiv - { 1\ov 12}
  \bigg[ (F^{s_1} ... F^{s_6}) - { 3 \ov 8 } (F^{s_1}... F^{s_4}) 
(F^{s_5}  F^{s_6})
+ { 1 \ov 32}   (F^{s_1}  F^{s_2})(F^{s_3}  F^{s_4})(F^{s_5}  F^{s_6})
      \bigg] \ ,  
\la{sixxx}
\ee
where 
$(F...F)$   indicates traces of  matrix products over
 suppressed Lorentz indices, 
 and 
$$
    \SSTr (Y_{s_1}  ... Y_{s_6} )  \equiv
2N \tr [Y_{(s_1}  ... Y_{s_6)} ]
 + \a_1  \tr [Y_{(s_1}  ... Y_{s_4}] \tr [Y_{s_5} Y_{s_6)}]
 + \a_2 \tr [Y_{(s_1}  ... Y_{s_3}] \tr [ Y_{s_4}... Y_{s_6)}]
$$ 
\be 
+\ \a_3 N\inv  \tr [Y_{(s_1}  Y_{s_2}] \tr [ Y_{s_3} Y_{s_4}]
\tr [ Y_{s_5} Y_{s_6)}]
 \ .  
\la{yytr}
\ee
$\SSTr (Y_{s_1}  ... Y_{s_6} )$
  is equal to $\STr (Y_{s_1}  ... Y_{s_6} ) $, 
when    
$\a_1= 30, \ \a_2= -20, \ \a_3=0$ (cf. \rf{rtr}).
 The condition  (implied by the  SYM--supergravity correspondence 
in the $0-(p+...+0)$ and $0-(1\pa 0)$ cases as  mentioned above) 
that $\hat C_6$ should  coincide with 
$\STr C_6$ \rf{eee} for the simplest commuting backgrounds  
with all $F_{ab}$ components proportional to the same $su(N)$ matrix, 
gives the constraints
\be
\a_1 = 30- \a_3 \ , \ \ \ \   \a_2 = \a_3 - 20  \ .  
\la{rree}
\ee
 Then \rf{yytr} becomes (cf. \rf{eeer},\rf{oonn}) 
$$
\SSTr (Y_{s_1}  ... Y_{s_6} ) = \STr (Y_{s_1}  ... Y_{s_6} )
$$
$$
 + \  \a_3 \bigg( -   \tr [Y_{(s_1}  ... Y_{s_4}] \tr [Y_{s_5} Y_{s_6)}]
 +  \tr [Y_{(s_1}  ... Y_{s_3}] \tr [ Y_{s_4}... Y_{s_6)}]
$$
\be
+\   N\inv  \tr [Y_{(s_1}  Y_{s_2}] \tr [ Y_{s_3} Y_{s_4}]
\tr [ Y_{s_5} Y_{s_6)}]\bigg) 
\ . 
\la{rrr}
\ee
We will  show that  demanding  also the 
agreement between the subleading term $\V^{(2)}$
in the supergravity potential \rf{vvv}  for the $0 - (4\pa 0)$ system 
and the 2-loop term in $\G$ for the corresponding (instanton)  
 gauge field  background  implies that 
\be 
\a_3 = -30 \ .  
\la{ttee}
\ee
To summarise,  
  the expression \rf{eeer} for $\hat C_{2n}$  and 
 the definition  \rf{see} of the polynomials $C_{2n} (F)$
lead to   the following all-loop BI-type ansatz  for the 
relevant part $\G$ \rf{eff}  of the  SYM effective action 
\be
\G=  { 1 \ov 2 N \gym^2} \int d^{p+1} \td x  \
 \SSTr \bigg[ H^{-1}_p \big(\sqrt{-\det{\big(  \eta_{ab} I  +  H^{1/2}_p  F_{ab} }\big)}
- 1 \big)  + { \four } F^2 \bigg] + ...  \ , 
\la{ohhh}
\ee
\be  H_p \equiv { a_p N \gym^2 \ov M^{7-p}} \ .   
\la{hre}
\ee
The dots in \rf{ohhh} stand for  possible  commutator terms
which  vanish  for  
commuting $F_{ab}$.   
$I$ is the unit  $N\times N$ matrix and  the square root of the 
 determinant is understood
as in the  expansion  \rf{see} with  each $F_{ab}$  now  replaced 
by an $su(N)$ 
matrix.\foot{Note that in contrast to the classical 
non-abelian BI Lagrangian  \rf{born} 
\ci{dbi}, i.e.
 $L \sim g_s\inv {\rm Str}  \sqrt{-\det{\big(\eta_{ab} I +  F_{ab} \big )}}$
  which represents a part of 
 the tree-level open string effective action and  is defined in terms of the 
 symmetrised trace in the  fundamental
representation,  $\G$, 
  which is a sum  of  certain   quantum loop  corrections, 
is defined in terms of the modified
symmetrised trace  $\SSTr$.
 Expressed in terms of  the traces in the fundamental representation 
$\SSTr Y^n$  starts with the $2N{\rm Str} Y^n $  term 
but contains also terms with multiple  traces $\tr$ (see \rf{trac}).}

Let us finally   turn to the   general 
case of {\it non-commuting}
$F_{ab}$ backgrounds for which  additional  commutator terms which, in general, may be present in  
$\hat C_{2n}$  with $n >2$
  may   be non-vanishing.
An  example of  a background with $[F_{ab}, F_{cd}]\not=0$
will be  used  in section 5.2 to describe the  1/8 supersymmetric bound state $4\bot4\bot4\pa 0$.
 We   shall  assume that  the 2-loop coefficient $\hat C_6$
is given by the totally symmetric expression
$\SSTr C_6 (F)$   \rf{oonn} plus  an order $F^6$  commutator term $\C_6$
\be
 \hat  C_{6} (F) =  
 \SSTr \ C_{6} (F) +  \C_{6}(F) \  ,  
\la{eeerrr}
\ee
\be
 \C_{6} \sim  \Tr (F F [F,F] FF)  + ...  \ .  
\la{sixe}
\ee
In general,  $\C_6$ may have  different Lorentz index structure than $C_6$
 \rf{sixx} and different internal index structure than 
a single $\Tr$.  
Though we  will not be able to determine  the detailed form 
of $\C_6$, we 
will  see that the presence of such commutator terms
is  necessary for the  complete 
correspondence between the supergravity and SYM expressions for the 
subleading term in the interaction potential of
 a 0-brane with 
the $4\bot 4 \bot 4 \pa 0$  bound state.
The  single $\Tr$ form of  the representative term in $\C_6$ in \rf{sixe} 
  will be  important for  getting  the  correct  $n_0 N^2_0$  scaling 
of the  whole 2-loop correction.

\newsection{$0$-brane  interaction with 1/2 supersymmetric D-brane bound states}
Below we shall consider the interaction potential between 
a 0-brane  and  a non-marginal 1/2 supersymmetric
bound state $p+ (p-2)+ ... +0$ of type IIA 
D-branes ($p=0,2,4,6$). Special cases  are 
the 0-brane -- 0-brane  and 0-brane -- $(2+0)$-brane
interactions. From M-theory point of view  this corresponds to  scattering of a 
$D=11$ graviton  (with fixed finite $p_-$ or fixed large $p_{11}$) 
off a  transverse  `p-brane'.

We shall demonstrate that the  classical 
supergravity  expression for the potential $\V$  \rf{vvv}  admits a 
 SYM interpretation (to all orders in string 
coupling)  which is equivalent to   a special case of  \rf{eff},\rf{ohhh}. 
In particular, the $v^6/r^{14}$ term in the  
0-brane -- 0-brane  interaction \ci{bbpt} and  the 
corresponding  terms in  the more  general  $0 - (p+...+0)$ case originate
from  the  same  2-loop $F^6$-term in  \rf{eff}. 
This generalises   the  previously known  relation between  the leading-order
term in $\V$ and  the 
 $F^4$ term in the  1-loop SYM effective action \rf{hh}. 

In the case of  the scattering off 1/2 supersymmetric bound states of D-branes the 
correspondence with the SYM theory is  rather explicit 
   because  of  the simple BI structure of the  action of the  $(p+...+0)$
brane  treated as a probe. 
 This  correspondence 
will no longer be so  transparent for the 0-brane interactions with 
 1/4  and 1/8  supersymmetric  bound states discussed in 
 sections 4 and 5; 
these cases will  provide  non-trivial checks
of the  consistency of the ansatz  \rf{eff},\rf{eeer},\rf{ohhh}.

\subsection{Supergravity expression for the $0-(p+...+0)$ potential}
To describe the interaction of a 0-brane and  a type IIA 
$(p+...+0)$ brane   we shall follow \ci{CT1,CT2} and consider the action of  $(p+...+0)$ brane  (i.e. a Dp-brane action with a constant abelian background $\F_{mn}$) as a probe
moving in the  background produced by a 0-brane as a source.
One can check 
that  the same expression  (to {\it all} orders in velocity and charges)  is found by considering a 0-brane as a probe moving in a background 
produced by the $(p+...+0)$ brane as a source. 
The $(p+...+0)$ probe 
action  is (see \ci{CT1,CT2}) 
\be
I_p = - T_p  V_p \int dt \ \bigg[ H_0\inv  \sqrt { (1 - H_0 v^2) \ \det( { H_0^{1/2} \delta_{mn} 
+ \F_{mn} }) }  - H_0\inv \sqrt{ \det\ {  \F_{mn} }} \bigg]  \ . 
\la{one}
\ee
We have assumed that the coordinates $X_i$
 transverse to the  $p$-brane depend only on the world-volume time $t$
with $v$ being  the velocity in a transverse direction. 
 $V_p$ is the volume of a torus  around which 
$p$-brane is wrapped. 
The Dp-brane tension  $T_p$ is \ci{poll}
\be
  T_p \equiv n_p \bar T_p=  n_p g_s\inv (2\pi)^{(1-p)/2}  T^{(p+1)/2} \ , \ \ \ \ \ \
   T\equiv (2\pi \a')^{-1} \ ,  
  \la{two}
  \ee
where  the integer $n_p$ is the number of p-branes (in what follows $T=1$). The  0-brane charge   of $(p+...+0)$ brane  is
\be
n_0 = n_p  (2\pi)^{-p/2} V_p  \sqrt{\det\ \F_{mn}} \ . 
\la{three}
\ee
$H_0$ is the harmonic function corresponding to the 0-brane  source
\be
H_0 ={Q_0^{(p)}\ov r^{7-p}} \ ,  
\la{four}
\ee
where $Q_q^{(n)}$ denotes the `charge' of  a Dq-brane 
background smeared in $n$ dimensions, 
\be
Q^{(n)}_q = N_q g_s (2\pi)^{(5-q)/2} T^{(q-7)/2} (V_n \om_{6-q-n})\inv
\ , \ \ \  \ \ \ \
 \om_{k-1} = {2 \pi^{k/2}\ov \Gamma({ k\ov 2})}  \  . 
\la{fif}
\ee
As already noted in the Introduction, we shall 
 drop  out the usual asymptotic value $1$ in $H_0$
(assuming either   the `null reduction' or `fixed  $p_-$' prescription 
  \ci{bbpt}  or 
that the 0-brane charge   $Q_0^{(p)}  \sim  V_p\inv g_s N_0$   is large, 
 so that
 ${Q_0^{(p)}\ov r^{7-p}} \gg 1$).
It is under  this prescription 
 that the expression for the resulting interaction potential
will have a simple SYM interpretation.

The action \rf{one}
may be rewritten as 
\be
I_p = - T_0  
\int dt \  H_0\inv  \bigg[ \sqrt { 1 - H_0 v^2} \sqrt{
\det ( { \delta_{mn}  + H_0^{1/2}\tF_{mn} }) }  - 1 \bigg]  \ ,  
\la{six}
\ee
where $T_0 = n_0 g_s\inv (2\pi)^{1/2}$ and 
\be \tF_{mn} \equiv (\F_{nm})\inv  \  .   \la{jjjj}  \ee
This corresponds to a T-dual  configuration   obtained 
by applying   T-duality
along  all of the directions of the $p$-torus, i.e.
\rf{six} describes  the interaction of a $p$-brane source (with charge $N_0$) 
with  parallel 
$(0+...+p)$-brane probe  (with  $p$-brane charge $n_0$) 
 moving in a relative transverse direction.

As suggested  by \rf{six}, 
a  gauge theory description   should be based on  a   SYM  theory 
in $p+1$ dimensions.
It is natural, however, to  go one dimension higher 
 to  be able treat  the transverse velocity on the same formal 
footing with  
the  gauge field components $ \tF_{mn}$.
Let us assume that the direction of motion is $X_9$. 
T-duality along this direction transforms
the configuration in question into a static system
of a D-string parallel to $(p+1) + ...+ 1$ brane
with velocity becoming an electric field background.
Indeed, \rf{six} is equivalent to 
\be
I_p = - T_0 \int dt \  H_0\inv  \bigg[ \sqrt { -
\det ( { \eta_{ab}  + H_0^{1/2}\tF_{ab} }) }  - 1 \bigg]  \ ,  
\la{ssix}
\ee
where $a,b=0,1,...,p,9$  and 
$$ \tF_{09}=v \ . $$
In this   form  the action \rf{six} is   the same as for the  type IIB 
  T-dual configuration  of a D-string interacting with  a
$(p+1) + ... + 1$ brane or  for 
  a D-instanton  interacting 
with a  $( p-1)+ ... + (-1)$ brane (the relation between the 
0-brane and instanton cases  involves  
duality  in the euclidean time direction). 
The action of the  composite brane  probe
in the instanton background  is \ci{CT2}
(we are ignoring the dependence  of the brane coordinates 
 $X_i$ on  the world-volume coordinates)
$$
I_{p-1} = - T_{p-1} V_{p} 
 H^{-1}_0 \bigg[  \sqrt { \det ( H^{1/2}_0 \delta_{ab} 
 + \F_{ab}) } - \sqrt {  \det \  \F_{ab}} \bigg]
$$
\be
 = - T_{p-1} V_{p} 
\sqrt {  \det \ \F_{ab}}
 \  H^{-1}_0  \bigg[ \sqrt { \det ( \delta_{ab} +  H^{1/2}_0 \F^{-1}_{ab}) } - 1\bigg]  \ .  
\la{acc}
\ee
Here $H_0$  has the interpretation of 
the harmonic function of the instanton background (smeared in the 9-th direction).
This expression is indeed equivalent to \rf{six} for constant $v$ and $\F_{mn}$.

The interaction potential $\V$  in  \rf{ssix} written as 
\be 
I = \int dt\ \big( \ha T_0  v^2 - \V  \big) \ ,  
\la{poo}
\ee
is thus 
\be
 \V  =
 T_0 \sum_{L=1}^{\infty} C_{2L+2}(\tF) \ H_0^L \ ,  
\la{pot}
\ee
where $C_{2n}(\tF) $ are the polynomials of degree $2n$ 
in $\tF_{ab}$
 which appear in  the expansion   
of the BI action \rf{see}.
The explicit form of the potential is thus
found to be (see  (3.2),(3.5))
\be
\V =\sum_{L=1}^{\infty} \V^{(L)} =
  n_0 N_0 \td V_p\sum_{L=1}^{\infty}  \bigg( \frac{a_p }{r^{7-p}}\bigg)^L  (g^{2}_{\rm YM} N_0)^{L-1}\ 
C_{2L+2}(\td \F)  \ ,  
\ \
\label{vl}
\ee 
where  
$ a_p = 2^{2-p}\pi^{-(p+1)/2}
{\Gamma ({\textstyle{ 7-p \ov 2}})}$
and $\td V_p$ is the volume of the dual torus, 
$V_p \td V_p = (2\pi)^{p}$.
We  have defined 
  $g_{\rm YM} =[(2\pi)^{-1/2} g_s \td V_p]^{1/2} $  as 
the effective coupling of the corresponding SYM
theory \rf{coe}.

\subsection{Relation to SYM effective action}
The dependence  of $\V$ \rf{vl} on the string coupling $g_s$ or on $ \gym^2$ 
suggests  that the  $1/r^{(7-p)L}$ term in it  should originate from 
the $L$-th loop  contribution   in  the relevant part $\G$ 
of the SYM  effective action on the dual torus.
As we  shall  demonstrate, 
 \rf{vl}  is indeed   reproduced 
by our ansatz  
   \rf{eff},\rf{ohhh}  for $\G$  to {\it all}
orders in the long-distance expansion.  
 For the  simple 
gauge field background  which describes the  $0-(p+...+0)$ configuration 
$\SSTr\ C_{2n} = \STr\ C_{2n}=\Tr C_{2n}$, i.e. the correction term in \rf{eeer}  vanishes
and $\hat C_{2n}(F)$ in \rf{eff} is given by \rf{eee}.
The correspondence  between $\G$ and $\V$ 
was previously  checked \ci{lifmat,lif3,CT1,CT2} at the 1-loop level 
  where $\G$ is given by \rf{hhh}.\foot{Note that the structure of the 
 $\hat C_6$  (2-loop) term in \rf{eff} is {\it different}  from the 
 $O(F^6)$ term in the 1-loop  SYM effective action (even though 
the latter  also has  a single $\Tr$ structure); in particular, 
as can be seen from the  general expression for the 1-loop effective action in a constant abelian background in \ci{CT2} or from  the results of 
 \ci{bachas,dkps}, 
the 1-loop $O(F^6)$ term  vanishes for the abelian $D=2$ 
background (and thus does not contribute to the 0-brane scattering)
while the 2-loop $F^6$ term does not. 
The vanishing of the coefficient of the  1-loop $v^6/r^{10}$  term in 
the  SYM expression for the 0-brane interaction potential 
was noted  in \ci{bbpt}.}

To compare \rf{vl} to \rf{eff} we need to identify the corresponding
 $u(n_0 + N_0)$  SYM  background which should be substituted  into $\G$  \rf{eff}.
It is given by \rf{baccc},\rf{becc} with  $A_m$ having constant field strength
given by $\tF_{mn}$ times  the  unit matrix $I_{N_0 \times N_0}$.
As  was noted in section 2, $\G$ and thus $\hat C_{2n}$ should, in general, depend
also on the scalar field background $\hat X_i$.   
To determine the dependence of $\G$ on the velocity $\del_0 \hat X_9$
one may use the 
the fact the lower-dimensional SYM 
theory  is a dimensional reduction of $D=10$ SYM theory which contains
 only  the gauge  field.  Since  $\hat C_{2n} (F)$  
are universal functions
of $F$ which do not explicitly depend on a space-time dimension, 
one may  expect that the dependence on $D_a X_i$  should 
 be the same as 
on $F_{ai}$  in  the coefficient 
 $\hat C_{2n} (F)$  in  a  higher-dimensional SYM theory.
Though this is not true in general 
for  the full effective action in SYM 
theory  (cf.  \ci{fraa}), 
 this  should be true for the part 
of the effective action $\G$ \rf{eff} we are interested in, since it  
 should  originate
from the open string loop diagrams with boundaries on the 
two different D-branes 
and thus should be 
covariant under the  T-duality \ci{poll, bachas} interchanging the abelian 
$A_k$ and $X_k$ components.
One   then  needs to compute $\hat C_{2n}$
in  a higher-dimensional SYM with  $F_{ai}\to D_a X_i$. Note, however, 
that does not mean that the full $\G$ is computed in a higher-dimensional space:  
the structure of \rf{eff}, i.e. the power of  $M$ and the 
 definition of $\gym$
are the same as in the original $D=p+1$  dimensional SYM theory with a  scalar
field background.\foot{A somewhat 
different  description  of the 1-loop result which is closely related to
the discussion  of the  D-instanton -- Dp-brane interactions in \ci{CT2}
is based on adding one extra 9-th dimension, choosing time to be euclidean  and compact 
and assuming that the 2-space $(\td x_0,\td x_9)$ has volume $V_2$ 
related to the  velocity as 
$ f_0 V_2= {2\pi}, \ f_0=iv .$
As in \cite{CT2}  one can show that for the background \rf{tra}
the  $F^4$-term in \rf{bbb} becomes 
$
{\bf b}_8=2 n_0[N_0f_0^4-f^2_0{\rm tr}(F_{mn}F_{mn})] \ , 
$
so that the 1-loop effective action is ($D= p+1 \to D+1$)
$$
\Gamma = i\frac{n_0}{2(4\pi )^{\frac{p+1}{2}}}\int_0^{\infty}ds
s^{\frac{6-D}{2}} e^{-sr^2}\int d^{p}x
\ \big[v^3N_0+\ v\ {\rm tr}(F_{mn}F_{mn})\big]  
= - i\int_{-\infty}^{\infty}d\tau\  \V^{(1)}  (r\to \sqrt{r^2+v^2\tau^2}) \ , $$
where 
$
\V^{(1)} (r)  =-\frac{n_0}{(4\pi )^{\frac{p+1}{2}}r^{7-p}}\Gamma({\textstyle \frac{7-p}{2}}) 
\int d^px \big[v^4N_0+\ v^2\ {\rm tr}(F_{mn}F_{mn})\big].$
}

The  basic example is  the  interaction 
between two  parallel p-branes one of which has a  velocity in the  direction 9
transverse to the  world volume of a p-brane  described by 
the  scalar  field  background
  $\hat X_9 $  \rf{becc} in $D=p+1$ dimensional SYM theory.
To determine the dependence of $\G$ on $\del_0 X_9$
one  is  thus  to compute $\hat C_{2n} (F)$  in $D'= D+1$ dimensions 
 in the electric background  $F_{09} = \del_0 X_9 \sim v $
and substitute  the result into  the original expression \rf{eff}
for $\G$ in 
$D=p+1$ dimensions.
Similar considerations
apply in  the case of oscillating 
 $X_i$ field  representing  a 
 wave carrying momentum  in some direction along the brane (for example,  
$X_1 = X_1 (x_5 +t)$ is related to a  gauge field 
wave $ A_1 = A_1(x_5 + t)$  in one higher dimension, see section 4.1 below).

The   pure gauge field    background 
in the `auxiliary' $D'=p+2$ dimensional SYM theory  which  corresponds to  
 the scattering of a  0-brane  
off a  $(p+...+0)$  bound state  is thus 
represented by the following 
gauge field matrices in the fundamental representation 
of $u(N )$\  ($N= n_0 + N_0)$
\be
\hat F_{09} =  \pmatrix{   v I_{n_0 \times n_0} & 0 \cr
  0  &    0_{N_0 \times N_0}\cr} \ , \ \ \ \  \ \ \ \ 
\hat F_{mn} =  \pmatrix{  0_{n_0 \times n_0} & 0 \cr
  0  &   - \tF_{mn} I_{N_0 \times N_0}\cr} \ .  
\la{tttr}
\ee
It  is useful to subtract the traces and  to describe  the background 
by the  $su(N)$  matrices $F_{ab}$   which are proportional to the same 
 matrix $J_0$ 
\be
F_{09} = \tF_{09} J_0  = v J_0 \ ,  \ \ \ \  \ \ \ \ 
F_{mn} =   \tF_{mn} J_0  \  ,   
\la{tra}
\ee
\be 
 J_0\equiv {1 \over n_0 + N_0  } 
 \pmatrix{  N_0  I_{n_0 \times n_0} & 0 \cr
  0  &   - n_0  I_{N_0 \times N_0}\cr} \  ,  \ \ \ \ \tr J_0=0 \ .  
\la{jjj}
\ee
 Since all of the components  of $F_{ab}$ commute
and
 $\Tr J_0^{2n} =  2 n_0 N_0$ 
one finds that\foot{Given a diagonal matrix in the fundamental representation
of $u(N)$  with entries $a_i$ 
the corresponding matrix in the adjoint representation
has  entries $a_i-a_j$.
That implies that $J_0$ has $ 2n_0 N_0 $ non-vanishing 
diagonal  elements
equal to $\pm 1$.}
 \be
\STr [ C_{2L+2}(F) ]  =  \Tr [ C_{2L+2}(F) ]   = 2 n_0 N_0 C_{2L+2}(\tF) \ .
\la{rrry}
\ee
To reproduce
\rf{vl}  one  should take
 only the $n_0 N_0^{L} v^{2L+2}$ part   of the $L$-loop 
term in \rf{eff},\rf{ohhh}  since the supergravity calculation 
based on `probe-source' picture is  asymmetric in $n_0, N_0$, i.e. 
gives   only  the terms which are  linear in the probe charge $n_0$.\foot{Equivalently, this corresponds to assuming  that the source is much heavier than the probe, i.e.
$N_0 \gg n_0$,  so that one may  replace $N=n_0 + N_0$ by $N_0$.} 
As a result, $\V$  \rf{vl} is in  precise agreement   with 
 $\G$ \rf{eff},\rf{eee}.

It is important to stress that 
 the presence of only  a single $\Tr$ 
in  the ansatz \rf{eff},\rf{eee}  is crucial for the correspondence between $\V$ 
and $\G$:  this trace produces 
 the $n_0 N_0$ factor, while the 
additional powers of  $N=N_0 + n_0$  
 are correlated with the power of the 
gauge field coupling  as implied by \rf{yyxx},\rf{stru}.

Demanding  that $\hat C_{2n} = \SSTr C_{2n} = 2 n_0 N_0 C_{2L+2}(\tF)$
or $\SSTr J_0^{2n} = 2 n_0 N_0$ implies that the correction terms
in \rf{eeer} must vanish.
In particular, starting with the most general 2-loop combination 
\rf{yytr} and  requiring that $\SSTr J^6_0= 2 n_0 N_0 $
fixes the two coefficients $\a_1$ and $\a_2$  according to \rf{rree}, 
i.e. the correction term in \rf{rrr} vanishes for $Y_s = J_0$.

\subsection{Special cases: $0-0$, \ $0-(2+0)$ and $0-(4+2+0)$}
The simplest special case is that of the  0-brane -- 0-brane  scattering.
Here $p=0$ and $\del_0 X_9 \to F_{09} =  v  J_0$ so that 
the  $1+0$ dimensional 
SYM effective action \rf{eff} in its closed  BI-type form \rf{ohhh}
becomes 
\be
\G=  {  n_0 N_0 \ov  N \gym^2} \int d t 
 \bigg[ H^{-1}_0 (\sqrt{ 1-   H_0 v^2 }
- 1)   +  \ha v^2  \bigg]  \ .
\la{ohh}
\ee
The factor $2n_0N_0$  came from $\Tr J_0^{2k}$ and  ($a_0 = {15\ov 2}$, $M=r$)
\be  H_0 =  { Q_0 \ov r^{7}} \ , \ \ \ \ \ \ \ Q_0=  
 { 15 \over 2(2\pi)^{ 1/2}}  N_0 g_s  \  .  \la{fooo}
\ee
This  agrees with the  exact expression for the 
  supergravity 
potential following from the 0-brane probe 
action 
$ I_0 = - T_0 \int dt H_0^{-1} \sqrt { 1 - H_0 v^2}$
after we  use  that  $T_0= n g^{-1}_s (2\pi)^{ 1/2}$ 
(see \rf{two},\rf{fif})
  and replace $N=n_0 + N_0$ in \rf{ohh} by $N_0$,  
i.e. separate  the $n_0 N_0^{L} v^{2L+2}$ part \ci{bbpt} 
of the $L$-loop 
term in \rf{ohh}.
This is the extension  of  the previous one-loop  ($v^4/r^7$)  \ci{bachas,dkps,bfss,ber}
and  two-loop ($v^6/r^{14}$) \ci{bbpt} results to all orders
in $1/r^7$  or loop expansion.   
Similar conclusion is  reached  also 
in the case when some  $p$ of the spatial 
dimensions
are compactified, i.e. the $0-0$ system is described by a  $p +1$
dimensional SYM theory.

Our discussion  also  clarifies 
the SYM structure of the 2-loop result of \ci{bbpt}. 
The $v^6/r^{14}$ term  in $\V$  considered in 
\ci{bbpt} corresponds to the 2-loop  $\hat C_6 \sim \Tr F^6$
term in the SYM description.\foot{Let us 
stress again  that the single $\Tr$ form of $\hat C_6$
is crucial for getting the correct $n_0 N_0^2$ factor in this  first 
subleading  term in the 0-brane interaction potential 
(this would not be so for a general combination of 
single $\tr$ and double $\tr$ terms in  \rf{yytr}).}
The  general $L=2$\  $F^6$ term in \rf{eff} 
  generalises  the  $1+0$ dimensional SYM 
result of  \ci{bbpt} to a  higher-dimensional 
case. The explicit 2-loop 
computation of $\G(v,r)$ in \ci{BB,bbpt} provides the overall 
normalisation of the 2-loop term   chosen
 in our ansatz for  $\G$  \rf{eff},\rf{coee}.
 Thus the  checks of consistency   of the detailed
structure of the 2-loop part of $\G$    discussed   below
will be concerned only with the relative normalisations
of  different terms in $\hat C_6$. 


The probe action  for the  $0-(2+0)$  interaction 
{\ci{ab,lifmat,ballar,CT1} is a special case of \rf{one}\foot{The action of the same structure is found by considering a 0-brane probe moving in the background 
corresponding to the non-marginal 
$2+0$  bound state. Using the explicit form of the $2+0$ solution \ci{RT}
one finds  for the 0-brane action:
$I_0 = -T'_0 \int dt [  K\inv  \td K^{1/2} \sqrt { 1 - K v^2} - \cos \theta (K\inv -1 ) ]$, 
where  $K= 1 + Q_2'/r^5$, \ $\td K = \cos^2\theta + K \sin^2\theta$ and 
 $\tan \theta = f_1= n_0/n_2 $
 (we consider  the self-dual torus with $V_2 = 2 \pi$). This action 
becomes  equivalent to  the action  in \rf{acd}  
provided we make identifications $H_0=K$ and 
$T_0= T_0'\cos \theta$.  The condition  $H_0=K$ or $Q'_2=Q_2 \sqrt{1 + n^2_0/n_2^2} = Q_0^{(2)}$ is satisfied in the limit of large $n_0$
(the relation between tensions  is also satified in this limit). }
\be
I_2 =  - T_0 \int dt \ 
H_0\inv \bigg[\sqrt { (1 -H_0 v^2 ) ( 1 + H_0 f_1^2) } - 1 \bigg]  \ ,  
\la{acd}
\ee
where $f_1= (\F_{12})\inv = (2\pi)\inv V_2 n_2/n_0$.
Since in  the  general case of a  block-diagonal 
euclidean 
matrix $F_{ab}$ with non-zero  entries $f_k$ ($f_0=iv, \ f_1 = F_{12} , \ f_2=F_{34}, ...)$  the polynomials $C_{2n}$ in \rf{fff},\rf{sixx} are 
 \be C_4 = - { 1 \ov 8 } \bigg[  2 \sum_{k} f^4_k 
-  (\sum_{k} f^2_k)^2 \bigg]  \ ,  
\la{yyy}
\ee
\be 
C_6 =  {  1 \ov 12} \bigg[  2 \sum_{k} f^6_k 
 - { 3 \ov 2 } \sum_{k} f^4_k  \sum_{n} f^2_n 
+ { 1 \ov 4} (\sum_{k} f^2_k)^3 \bigg]  \  ,   
\la{too}
\ee
we conclude that 
the  leading term in the resulting 
potential $\V^{(1)} \sim  { 1 \ov r^5} (v^2 + f_1^2)^2$ 
originates from the one-loop term $\sim \Tr
 [F^4 - { 1\ov 4} (F^2)^2]$
 in \rf{eff}
while the first subleading term $\V^{(2)} \sim  { 1 \ov r^{10} }
(v^2 - f^2_1) (v^2 + f_1^2)^2$ 
is reproduced by the 2-loop SYM term  $\sim \Tr 
[F^6   - { 3 \ov 8 } F^4 F^2  
+ { 1 \ov 32} ( F^2)^3\bigg] $ with $F_{09}= v J_0, \ F_{12} = f_1 J_0$
as in \rf{tra}.
Note that since  we have  subtracted 1 in $H_0$ we do not  need to 
consider the limit of   large  field $\F_{mn}$
( i.e. of   large $n_0 \gg n_2$)  in order to establish the 
precise agreement 
between  the supergravity  and SYM expressions (cf. \ci{bbpt} and
 \ci{lifmat,ballar,CT1}). 

In the $0-(4+2+0)$ case  we  get
 one extra factor of $(1 + H_0 f^2_2)$
under the square root in \rf{acd}  ($f_2= (\F_{34})\inv$), so that
the leading and subleading terms in the interaction potential in \rf{vl} are  
\be
\V^{(1)} =  -{ 1 \ov 8 r^3} T_0 Q^{(4)}_0  \big[  (f^2_1 - f^2_2)^2 + 
2 v^2 (f^2_1 + f^2_2)     +  v^4  \big]     \ ,  
\la{sss}
\ee
 \be 
\V^{(2)} =  - { 1 \ov 16 r^{6} } T_0 (Q^{(4)}_0)^2 
(f^2_1 + f^2_2 + v^2)  [ - (f^2_1 - f^2_2)^2    + v^4]
    \ ,  
\la{soo}
\ee
and  are reproduced by the 1-loop and 2-loop terms in
 $\G$ \rf{eff},\rf{yyy},\rf{too}.
Note that in  case of the self-dual background $f_1 = f_2 = f$
the static terms in the  potential  cancel out 
(for $F_{mn}=F^*_{mn}$ the $D=4$ BI action 
becomes quadratic in  the field strength) so that 
\be
\V^{(1)}  = - { 1 \ov 8 r^3} T_0 Q^{(4)}_0  \big( 
4 v^2 f^4      +  v^4  \big)     \ , 
\ \  
\ \ \ \ \ 
\V^{(2)}  = - { 1 \ov 16r^{6} }T_0 (Q^{(4)}_0)^2 
\big(   2   v^4  f^2    + v^6 \big) \  .   
\la{sooo}
\ee

\newsection{$0$-brane  interaction with 1/4 supersymmetric
 bound states}
In this section we shall  study 
the subleading terms in the interaction potentials of a 0-brane probe 
with  1/4 
supersymmetric  marginal 
bound states of branes: a bound state of a  fundamental string and a 0-brane
$1 \pa 0$  and a bound state of a 4-brane and a 0-brane $4\pa 0$.
The same potentials 
describe  the  scattering of  $D=11$ gravitons (with fixed  
large $p_{11}$ or fixed finite $p_-$)  off  the bound states
of M2-brane  with wave and M5-brane  with wave 
(`longitudinal M2-brane' and `longitudinal M5-brane').
In contrast to the  scattering off 1/2 supersymmetric bound states
discussed in the previous section, 
here the supergravity expression for the potential
(again obtained from  a Born-Infeld-Nambu
 type  action  in curved space) is
 no longer  immediately  interpretable as  a  SYM  expression 
because of  a more  complicated
structure of the background fields (containing products
 of different harmonic functions).

This  structure  is not seen at the leading order 
in  large-distance expansion  since the leading term $\V^{(1)}$ 
in the potential depends  on the constituent charges in an `additive' way.
Indeed, $\V^{(1)}$ is reproduced by the  one-loop  ($\STr C_4 \sim \Tr F^4 + ...$)  term in  the SYM effective action (3.15).
The correspondence between  the supergravity and  SYM  (or matrix theory)
expressions for  $\V^{(1)}$ was previously established
in the  $0-(4\pa 0)$  case 
 in \ci{CT1} (with $F^4$ interpretation
given in \ci{CT2})  and   in the  $0-(1\pa 0)$  case  in \ci{GR}. 

We  will  find that  in  the $0-(1\pa 0)$  case
 the first {\it subleading} 
term  $\V^{(2)}$ in the potential 
 is  again  reproduced by 
the  2-loop term  in  $\G$ \rf{eff},\rf{eee},
 i.e.  it has the  same structure (and coefficient) as
 $\hat C_6 \sim \Tr F^6$.
Remarkably,  
this correspondence extends also to all higher-order terms
in $\V$ and $\G$ \rf{eff},\rf{ohhh},  just as in the
$0-(p+...+0)$   case of the previous section.
This is  a consequence of the fact that the corresponding YM 
background is  again proportional to a single $su(N)$ matrix.

The  situation in the $0-(4\pa 0)$ case  is 
 more complicated:
while the supergravity potential  has the same form as in the $1\pa 0$ case, the relevant  YM background is less trivial   as it now
 depends 
two  different (but still commuting)  $su(N)$ matrices. As a result, 
$ \V^{(2)}$ is reproduced by \rf{eff} with 
$\hat C_6$ given by \rf{eeer}, 
 i.e.  with  $F^6$ terms  having the same Lorentz-index structure 
but a modified  prescription  $\SSTr$ \rf{rrr} 
for taking traces over the internal  indices.  $\SSTr C_6$
 differs from $\STr C_6$  by the $\a_3$  term  in \rf{rrr} 
which  was 
vanishing  in all previous cases
 (i.e. $0-(p+...+0)$ and $0-(1\pa 0)$) but  turns out 
to be  non-zero in the $0-(4\pa 0)$  case.

\subsection{ $0$-brane  -- $(1\pa 0)$-brane interaction}
To find the interaction potential we shall consider 
a 0-brane probe moving in the background produced by  $1\pa 0$ as a source.
The  type IIA supergravity  solution   representing  the 
1/4 supersymmetric marginal 
bound state   of  a fundamental string and  a 0-brane 
is a  dimensional  reduction  of  the $D=11$ 
`2-brane + wave'  solution \ci{ts2,ts4}  and is given by 
\be
ds^2_{10} = H_0^{-1/2} \HH_1^{-1}[ -  dt^2
 + H_0  dx^2_5  +  H_0 \HH_1 dx_i dx_i] \ ,  
\la{back}
\ee
$$e^{2\phi} = \HH_1^{-1} H_0^{3/2} \ , \ \ \ \ 
A= H_0^{-1}dt \ , \ \ \ \  $$
\be
H_0 = {Q_0^{(1)}\ov r^{6}}\ , \ \ \ \HH_1 = 1 + {\Q_1\ov r^6} \ , 
 \ \ \ \  \   \Q_1 = g_s Q_1  = (2\pi)^{-1/2} g_s L_5 { N_1 \ov N_0}  Q_0^{(1)} \ ,    
  \la{www} \ee 
 where $\Q_1$ is the fundamental string charge ($N_1$ is the  winding number)
 and
 $Q_1$ and $ Q_0^{(1)}$ are defined in \rf{fif}.
 $L_5$ is the length of the circular  direction  (chosen to be the 5-th one)
along  which the fundamental string is wound. 

The  action of a  0-brane probe  with a transverse velocity $v$
 is then
\be
 I =- T_0 \int dt\  H_0^{-1} \bigg[\sqrt { 1 - H_0 \HH_1 v^2 } - 1 \bigg] \ , 
\la{acci}
\ee
so that  the first two terms in the  interaction  potential $\V$ \rf{pou},\rf{vvv}
 are thus    
\be
\V^{(1)} =  - { 1 \ov 8r^{6}}  T_0  ( 4 \Q_1 v^2 +  
  Q^{(1)}_0  v^4 )   \ , 
\ \ \ \ \  \ 
\V^{(2)} = - { 1 \ov  16r^{12}} 
 T_0 Q^{(1)}_0 (4 \Q_1 v^4  +    Q^{(1)}_0  v^6 ) \  .  
\la{vot}
\ee
It is the product of the  {\it two} harmonic functions 
under the square root in \rf{acci}
compared to a single   factor of $H_0$ in \rf{six},\rf{acd}
that makes comparison with a SYM action non-trivial.
Similar complex structure  of the probe action
containing the    product of {\it several} 
harmonic functions  \rf{pou}  is characteristic to 
all  cases of scattering off
1/4 and 1/8  supersymmetric  BPS configurations
discussed below. This structure is  a consequence of 
the `harmonic function rule'  form  of the supergravity backgrounds
representing  marginal   BPS  bound states of different
 branes \ci{ts2}, and  it 
 makes establishing a 
connection  between  a  curved space 0-brane action
 and a flat space SYM effective  action quite  non-trivial.

Since $T_0 \sim  g_s\inv n_0, \ \Q_1  \sim g^2_s N_1, \ 
Q^{(1)}_0 \sim  g_s N_0$ it may   seem  that $\V^{(1)}$ and
$\V^{(2)}$  contain terms of different orders in  the  string coupling.
Still,  
$\V^{(1)}$  is reproduced by the   1-loop  correction in the 
SYM  effective action, and 
$\V^{(2)}$  -- by  the 2-loop one, as  for  the pure D-brane configurations discussed in the previous section (and  for  the $4\pa 0$ case 
considered  below).  The reason is that after T-duality  the fundamental 
string winding number $ N_1$ will have  
the interpretation  of a  momentum carried by the classical SYM wave, 
 i.e. the  corresponding gauge field background
will explicitly depend on $g_s$, \ $F_{ab} \sim  {g_s^{1/2} }$. That 
 will bring in an extra power of $\gym^2 \sim g_s$  on the SYM side.

The $1\pa 0$ state  is, 
 indeed,  T-dual to a bound state of a D-string and a wave:
 the action  \rf{acci} 
 is the  same as for a D-string probe 
in  the D-string + wave background (with the  string probe oriented parallel
to the string source and moving in the  orthogonal direction). 
This  relation suggests that  the $0-(1\pa 0)$ 
 configuration should have a  description 
in terms of the $D=2$ SYM theory  
with $\Q_1$ having the  interpretation of a SYM  momentum 
\ci{imam,dvv,wu,gopak,GR}.
The latter is  represented by a  periodic scalar  field  
background 
$X_1= X_1(\td x_5+t)$  with $\td x_5$ being the  direction of the  momentum flow.

To find the dependence of the coefficients $\hat C_{2n}$ in 
the SYM effective action $\G$  \rf{eff} 
on derivatives of  $X_1$
we  may  
formally trade the $X_1$ wave  for a  gauge field 
 wave  by going to a  $D=3$  SYM theory 
and considering the T-dual  background $A_1= A_1(x_5+t)$.
This is  similar to the trick used  above
to find the  
 dependence on $\del_0 X_9\sim v$:
one  considers the   SYM theory  in one dimension  higher 
with an electric gauge field background $F_{09}$.

  Altogether this corresponds to performing 
T-duality along the direction of the  transverse D-string  momentum-carrying 
 oscillations ($x_1$) and  the  direction of motion of its center of mass 
  ($x_9$). We  end up with a configuration 
of two parallel  3-branes  described by a 
  a  plane  wave   (in the  direction  5)
 on one  3-brane  and a constant 
 electric field  (in the direction 9)   on  another 3-brane.
It  is represented  by the 
 the following stationary abelian  $u(N)$  gauge field background   in 
 the $D=4$  SYM theory  on the dual torus
 (cf. \rf{tra})
\be
\hat F_{09} =  \pmatrix{   v I_{n_0 \times n_0} & 0 \cr
  0  &    0_{N_0 \times N_0}\cr} \ , \ \ \ \ \ \ 
\hat F_{51} =  \hat F_{01} =
 \pmatrix{  0_{n_0 \times n_0} & 0 \cr
  0  &  - h(\td x_5+ t) \ I_{N_0 \times N_0}\cr} \ ,   
\la{trra}
\ee
or, equivalently,  by the  following $su(N)$ background 
\be
F_{09}  = v J_0 \ , \ \ \ \ 
\ \ \ \ 
F_{51} = F_{01} = h(\td x_5 +  t) J_0  \  ,  
\la{ttt}
\ee
where the $su(N)$ matrix  $J_0$  was defined  in \rf{jjj}.  The function $h$ 
which is  the derivative of  
 $A_1=A_1 (\td x_5 +t)$ 
may be chosen, e.g.,
 as $ h  \sim \sqrt { g_s}\sin  [{ 2 \pi \ov \td L_5} ( \td x_5 
 + t) ]$  \ci{GR} and  is normalised so that\foot{Let us note 
that for the aim 
of reproducing the expressions for interaction potentials  
 here and in all  examples discussed below
one, in principle,  does not need to know the  explicit form of the gauge
 field backgrounds  representing a plane wave or instanton or their 
superpositions -- all what is needed are the  basic properties like
constraints on the field strengths (i.e. $F_{1+}=0$  or $F_{mn} = F^*_{mn}$)  and  normalisation conditions for the integrals of the 
squares of the  field strength components.} 
\be
 < h^2 > =  { 1 \ov \td L_5 } \int d \td x_5  \ h^2  =  g_s  (2\pi)^{1/2}  { N_1   \ov N_0 \td L_5  }
= { \Q_1 \ov Q^{(1)}_0}   \ .  
\la{noo}
\ee
  $N_1$ is  thus  the momentum carried by the gauge field wave
along the  5-th direction, 
\be
{ 1 \ov (\gym^2)_3 } 
\int d \td x_1  d \td x_5 \tr (\hat F_{01} \hat F_{51})
 =  { 2 \pi N_1 \ov \td L_5}  \ , \ \ \ \ (\gym^2)_3 = (2\pi)^{-1/2} g_s \td L_1 \td L_5  \ .   
\la{norm}
\ee
Here  the dual torus dimension $\td x_1$  (with length 
$\td L_1$)  is  the  auxiliary 
  dimension of the  plane wave and $ \td L_5 $
is the length of the dimension $\td x_5$ 
dual to the one  along the fundamental string 
($ \td L_5 L_5  = 2 \pi$). 

It should be stressed  again that the passage to the  $2+1$ or $3+1$
 dimensional SYM theory
serves only to determine the dependence  on the
 derivatives of the scalar fields $X_1$ and $X_9$ in the original $D=1+1$ dimensional  SYM theory: 
to find  the  SYM  effective action $\G$ which should correspond to the 
supergravity potential $\V$  we should set 
$p=1, \ D=2, \ \gym^2 =(\gym^2)_2= (2\pi)^{-1/2} g_s \td L_5$ in \rf{eff}.

Since  all of the  components  of 
$F_{ab}$  in \rf{ttt} are, as in \rf{tra}, 
  proportional to the same $su(N)$ matrix $ J_0 $, 
the  expression for $\hat C_{2n}(F)$ in \rf{eff} is again given by 
\rf{eee} and,   as in \rf{rrry}, 
$\hat C_{2n} = 2 n_0 N_0 C_{2n} (F)$. Thus  we 
 only  need to compute  the coefficients 
 $C_4\sim F^4$ and $C_6\sim F^6$  in \rf{fff},\rf{sixx}   for the  abelian 
background $F_{09} =v , \ F_{51} = F_{01} = h$, 
\be
C_4 =  - { 1 \ov 8} ( 4 h^2 v^2 + v^4 )  \  , \ \ \ \ \ 
C_6 = - { 1 \ov 16}  ( 4 h^2 v^4 + v^6)  \ .  
\la{resu}
\ee
Using \rf{noo}\foot{The extra power of $\gym^2\sim g_s$  in the classical momentum \rf{norm} explains  the correspondence with
 the supergravity expressions in \rf{vot}
(see also \ci{dps,malda}).} we 
 find  the precise  agreement  between the two leading terms
in $\V$ \rf{vot} and the 1-loop
 and 2-loop terms in the SYM effective 
action \rf{eff}.\foot{As was already mentioned above, 
here and in  the examples discussed in the following sections, 
it is sufficient to check only the agreement 
between the supergravity and the SYM expressions 
for the relative coefficients between the highest power of $v$ 
and its lower powers at each order of $1/r^{7-p}$ expansion
 since 
the agreement of the  coefficients of the $v^4$ and $v^6$ terms
(same as in the 0-brane--0-brane scattering 
in the case of compactification on a torus)
was already established  by the  one-loop \ci{dkps,ber}
and two-loop \ci{BB,bbpt} computations.}  
 At the 
leading-order level this correspondence was  also
checked  by  direct 1-loop  $D=2$ SYM calculation  in  $X_1$ background in 
\ci{GR}.\foot{The potential  vanishes in the static limit
reflecting the  BPS nature of the plane wave background
(which preserves 1/2 of supersymmetry in SYM theory).
 Let us note in passing that the vanishing of the 1-loop YM effective action in the  non-abelian plane wave  background was discussed in \ci{arodz}.}


 As in the $0-(p+...+0)$ case,  the relation  
between  $\V$ in \rf{acci} and $\G$ in \rf{eff},\rf{eee},\rf{ohhh}
 holds  not only for the first two leading terms, but 
also for  the complete expressions, i.e. for all terms 
in the  expansion in $1/r^6$.
Indeed, $\G$ in \rf{ohhh} with $p=1$ and the $D=4$ BI determinant
$\det (\eta_{ab} I + F_{ab}) $  computed  on the abelian 
background \rf{ttt} 
(which looks the same as
 a D3-brane action in the 
 gauge field background or a D-string action 
in the  scalar field background)
  is found to be  (cf. \rf{ohh})
\be
\G=  { n_0 N_0  \ov  N \gym^2}
\int dt d\td x_5 \bigg[ H^{-1}_1 ( \sqrt { 1 - ( 1 + H_1 h^2) H_1  v^2 } 
 - 1)  + \ha v^2 \bigg] \ . 
\la{ohe}
\ee
Here $H_1$ given by  \rf{hre} is equivalent to $H_0= Q^{(1)}_0/r^6$ in \rf{www}.
This expression (its part linear in $n_0$) 
 takes exactly  the same form  
as $\V$ in \rf{acci},\rf{www} after we replace $h^2$ by $<h^2>$ in \rf{noo}
so that $1 + H_1 h^2$ becomes $\HH_1$ in \rf{www}. 
 This  establishes the  agreement between   all 
higher-order  terms in the expansion of the  interaction potential  in \rf{acci} and
the SYM effective action \rf{eff},\rf{eee}.

\subsection{$0$-brane -- ($4\pa 0$)-brane  interaction }
 The action of   a 0-brane probe  moving  in the background
produced by 
  the bound state $4\pa 0$ as a   source 
 is  \ci{ts3,CT1}
 \be
 I_0 = -T_0 \int dt \   H_0^{-1} \big( \sqrt {1- H_0 H_4 v^2  }
    -   1\big)  \ , 
 \la{zerr}
 \ee
 \be H_0= {Q^{(4)}_0\over r^3} \ , \ \ \ \ \ \  H_4= 1 + {Q_4\over r^3}
  \  ,  
 \la{rele}
 \ee
were  $Q_4$ and $Q^{(4)}_0$  are given by 
\rf{fif}. 
This action is formally the same as in the $0-(1\pa 0)$   case \rf{acci}
 with $H_4$  replacing  the  fundamental string  function $\HH_1$.
The  two leading terms in the  classical potential 
 thus have  the same form as in \rf{vot} with $ Q^{(1)}_0 \to  Q^{(4)}_0, \ 
\Q_1 \to Q_4$. 

The $4\pa 0$ brane  wrapped over a   4-torus $T^4$
and having even $N_0$  may be 
described by  the following self-dual  $u(N_0)$  background 
on the dual 4-torus $\td T^4$ \ci{doug,vafa,taylor,guraln}
 \be
F_{12}= F_{34}=
 q\ \sigma_3\otimes I_{{N_0\ov 2} \times {N_0\ov 2}}   
 \, ,   
\la{ggg} 
\ee
with all other components of $F_{mn}$ being zero. Here
$\sigma_3 ={\rm diag} (1, -1)$ and 
\be    q^2 =(2\pi)^2 \td V_4\inv  {N_4 \ov N_0} = 
{Q_4\ov Q^{(4)}_0 }  \ ,  \ \ \ { \rm i.e.}  \ \  \ \ 
{ 1 \ov 16 \pi^2 } \int_{\td T^4}  d^4 \td x  \ \tr (F_{mn} F_{mn}) = N_4 \ . 
\la {gggg}
\ee
The leading and subleading terms of the supergravity  potential,
expressed in terms of  the YM  background, read 
$$
\V^{(1)}
= -\frac{n_0}{16r^3} \bigg[4 v^2 N_4   + (2\pi)^{-2}\td V_4  v^4 N_0\bigg]
$$
\be
=-\frac{n_0 N_0 }{64\pi^2r^3}\td V_4  \bigg( 4 v^2 q^2 +v^4 \bigg) 
=-\frac{n_0}{64\pi^2r^3}\td V_4 
\bigg[v^2{\rm tr}(F_{mn}F_{mn})+v^4 N_0\bigg] \ ,  
\la{xxx}
\ee
$$
\V^{(2)}=-\frac{n_0N_0g_s}{64(2\pi )^{5/2} r^6} \td V_4 \bigg[
4 v^4  N_4   +  (2\pi)^{-2}\td V_4  v^6 N_0\bigg]
$$
\be 
= -\frac{n_0 N^2_0 g^2_{\rm YM}}{64(2\pi )^4r^6}
\td V_4 \bigg( 4 v^4 q^2   +  v^6\bigg)
= 
-\frac{n_0N_0g^2_{\rm YM}}{64(2\pi )^4r^6}\td V_4 \bigg[v^4{\rm
tr}(F_{mn}F_{mn})  +  v^6 N_0\bigg] \ . 
\label{vtwo}
\ee
While the leading-order potential $\V^{(1)}$  is the same
as in the special case of $0-(4+2+0)$  interaction  in \rf{sss}
with  the  $(4+2+0)$
brane  described by  the  self-dual abelian 
background $\tF_{12}=\tF_{34} = q $
(so that  \rf{xxx} can be found also by 
using  such $(4+2+0)$ configuration to represent  
$4\pa 0$  as a probe  and treating the 
0-brane as a source  \ci{CT1})
this is no longer so for the subleading term in the potential $\V^{(2)}$:
as follows from   \rf{too}  (with $f_0=iv, \ f_1=f_2 =q$) in the 
$0-(4+2+0)$ case  
 the coefficient of the `mixed' $v^4 q^2$ term  in \rf{sooo} 
is   {\it factor of  2 smaller} than  in the supergravity expression
\rf{vtwo}.
 This extra  factor of 2  in $\V^{(2)}$
is a  direct consequence  of the 
 different structure of  the action  \rf{zerr} 
 compared to   \rf{six}:   it originates from 
the coefficient 2 in $(H_4 )^2= 1 + 2 Q_4/r^3 + ...$
which appears in the expansion of \rf{zerr}.

The two  configurations, i.e.  $0-(4+2 +0)$ and $0-(4\pa 0)$, 
having  different subleading terms in the  supergravity potentials 
may still  be described by the same universal 2-loop
SYM action $\G^{(2)}$  because  the corresponding  $su(N)$
 gauge field backgrounds are {\it different}.
While  in  \rf{tra} all the components 
of the field strength are proportional to the same $su(N)$ matrix $J_0$, 
here the  electric (velocity)  component
$F_{09}$ and the instanton  (4-brane) components
$F_{12}=F_{34}$ are expressed in terms of 
 two  different 
 matrices $J_0$ and $J_1$ 
 from  the Cartan  subalgebra of $su(N)$, 
\be
F_{09} = v J_0 \ , \ \  \ \  \ \ \ \ \ 
F_{12}=F_{34} = q J_1 \ ,  
\la{joj}
\ee
\be
  J_1\equiv    \pmatrix{  0_{n_0 \times n_0}  & 0 \cr
 0  &  \sigma_3 \otimes  I_{{ N_0 \ov 2} \times {N_0\ov 2}}  \cr } =
 \pmatrix{  0_{n_0 \times n_0} & 0  & 0 \cr
 0  &   I_{{ N_0 \ov 2} \times {N_0\ov 2}}  & 0\cr
    0  &   0 & - I_{{ N_0 \ov 2} \times {N_0\ov 2}}\cr
                     } \  ,  \ \ \ \ \tr J_1=0 \ .  \la{jj}
\ee
 As a result,  this  background is  more 
sensitive  to the non-abelian structure of the 
coefficients  $ \hat C_{2n}$  \rf{eeer} in the SYM
effective action \rf{eff}.

Let us first recall  how  the 1-loop term in  \rf{eff}
reproduces \ci{CT1} the leading term in the potential \rf{xxx}.
Substituting the background \rf{joj} into $\hat C_4 = \STr C_4 (F)$ and using \rf{fff},\rf{yyy} we find that\foot{Since the $F_{ab}$
background here is commuting, $\STr C_{2n} = \Tr C_{2n}$.
  It is useful, however, 
to  keep symmetrisation in order to express $\Tr$-expressions in terms of 
$\tr$ ones.  
Note that the symmetrised product of 
 $\tr J^2\tr  J^2$  where $J= J_0$ or $J_1$ is 
 ${\rm Sym} (\tr J^2 \tr J^2)  \to  { 1\ov 3} 
[\tr J_0^2 \tr J^2_1 + 2 \tr (J_0 J_1) \tr (J_0 J_1) ] $. Thus
$\STr (J^2_0 J^2_1)  = 2 N \tr (J^2_0 J^2_1) + 2 \tr J_0^2 \tr J^2_1 = 2n_0 N_0$,
where  we have used \rf{traac}  and  $\tr (J_0 J_1) =0.$}
\be
\hat C_4  = \STr C_4 (F) = - {  1 \ov 8} \STr (4 v^2 q^2 J^2_0 J^2_1
+    v^4 J_0^4 )  = - { 1 \ov 4}  n_0 N_0  ( 4 v^2 q^2  + v^4) \ , 
\la{onee}
\ee
and, as a result,  complete agreement
  between \rf{eff} and  \rf{xxx}.\foot{Note again 
that since we have subtracted 1 from $H_0$ in \rf{rele} 
we do not need to assume (as was done  in \ci{CT1}) 
that $N_0 \gg N_4$ in the supergravity expression for the potential.}  

To reproduce the subleading term \rf{vtwo} in $\V$ we shall use 
$\hat C_6=\SSTr C_6 (F)$ in \rf{eeer}  with $\SSTr$ 
defined according to \rf{rrr}. 
Remarkably, the coefficient $\a_3$ in \rf{rrr} 
is fixed uniquely \rf{ttee} once 
 we    demand
the correspondence between $\V^{(2)}$ 
and the 2-loop SYM  effective action $\G^{(2)}$ in \rf{eff}.
The extra $\a_3$-term in \rf{rrr} is responsible for correcting the coefficient 2
of $v^4$ term in $\V^{(2)}$  in \rf{sooo} into 4 in \rf{vtwo}.
Indeed,  the  expression for $\hat C_6$ for 
 the background \rf{joj} is found to be (cf. \rf{too},\rf{sooo},\rf{onee})
\be
 \hat C_6 = - { 1\ov 16} \SSTr ( 2 v^4 q^2  J_0^4  J_1^2  + v^6 J^6_0)   \ .           \la{ghgh}
\ee
 Applying the definition \rf{rrr},\rf{ttee}  of $\SSTr$  
one finishes with\foot{Useful symmetrised 
products of traces of $J^4_0 J^2_1$  are 
  $
{\rm Sym} (\tr J^4  \tr J^2)  \to  { 1\ov 15} [\tr J_0^4 \tr J_1^2 +8 \tr (J_0^3J_1) \tr (J_0 J_1)  +6 \tr (J_0^2J_1^2) \tr J_0^2 ] \ ,  $ \ \ \ 
$
{\rm Sym} (\tr J^3 \tr J^3)   \to   {1\ov 5} [ 2 \tr J_0^3 \tr (J_0 J_1^2)
+3 \tr (J_0^2J_1) \tr (J_0^2 J_1 )] \  , 
$ \ \ \ 
$
{\rm Sym} (\tr J^2 \tr J^2 \tr J^2)  \to 
 {1\ov 5} [  \tr J_0^2 \tr J^2_0 \tr J_1^2+
4 \tr J_0^2 \tr (J_0 J_1) \tr (J_0 J_1 )] \  , 
$ \ \ 
which can be simplified using that $\tr J_0=\tr J_1= \tr( J_0 J_1) = 0$.
Let us note also that since, in general, 
$\tr (Y_{(s_1} Y_{s_2}) \tr (Y_{s_3} Y_{s_4}) \tr (Y_{s_5} Y_{s_6)})
= { 1 \ov 15} \big[ \tr (Y_{s_1} Y_{s_2}) \big( \tr (Y_{s_3} Y_{s_4}) \tr (Y_{s_5} Y_{s_6})  + 2 \ terms \big)  +  4\ terms \big]$, 
we find that 
$ {\rm Sym}( \tr X^2 \tr Y^2 \tr Z^2) \to 
{ 1 \ov 15} \big[ \tr X^2 [ \tr Y^2 \tr Z^2 + 2 \tr (YZ) \tr (YZ) ]  
   +      2 \tr (XY) [...        ]  + 2 \tr (XZ) [ ...]   \big]$. 
}
\be
 \hat C_6 = - { 1\ov 8N}   n_0 N_0  \bigg[ (n_0 + N_0 )  ( 
 2 v^4 q^2       + v^6 ) +   2  N_0  v^4 q^2 \bigg] 
   \ .  
\la{ete}
\ee
Isolating the part linear in $n_0$ in the 2-loop term in \rf{eff} which is 
proportional to $N \hat C_6$ 
we find the agreement with \rf{vtwo}.
Note that while in 
 the case of the 0-brane--0-brane scattering 
in \ci{bbpt} the 2-loop SYM expression for the potential
 was  probe--source ($n_0 \leftrightarrow N_0$) symmetric, 
this is no longer true  in the $0-(4\pa 0)$ case:
the symmetry is broken by the non-vanishing 
correction term in $\SSTr C_6$  producing the second $v^4 $ 
term in \rf{ete}.

\newsection{ $0$-brane  interaction with 1/8 supersymmetric
 bound states of branes}

By demanding  the precise agreement  between 
the supergravity potentials for the 
$0-(p+...+0)$, \ $0-(1\pa 0)$ and $0-(4\pa 0)$ cases
we have so far fixed the structure of the  2-loop term $\sim \hat C_6$
 \rf{yytr}--\rf{ttee}
 in the  leading IR part  $\G$ of the SYM effective action, at least
up to  commutator terms which vanish on commuting $F_{ab}$
backgrounds.  
A further  test of   consistency of our ansatz for $\G$ 
and thus  of  correspondence
 between  the supergravity and SYM  descriptions
is   provided by  non-trivial examples  of 
 $0$-brane  interactions with 1/8 supersymmetric
 bound states of branes. These bound states
are represented by intersections of three  and   four
 branes in $D=10$ (or $D=11$)  and, when wrapped over 5-torus and  6-torus,
are related to   $D=5$ and $D=4$ extremal black holes with 
regular horizons.

 We  will  show that these  bound state configurations
admit  simple SYM   descriptions (which, however, are not unique).
They are essentially  superpositions  of   plane  wave and/or instanton 
backgrounds  which  represent  1/4 supersymmetric BPS states
of SYM theory.\foot{Previous discussions of SYM/Matrix theory description
of $D=5$ black holes 
appeared in \ci{dps,limart,dvvv,halyo,malda};
some comments on $D=4$ black holes were also  made in \ci{limart}.}
Plugging these backgrounds into the $F^4$-term in the 1-loop
SYM effective action  demonstrates the agreement 
between  the supergravity and 
SYM expressions for the leading-order interaction 
potentials  (in the $D=5$ black hole 
case this was  previously   shown  in \ci{dps,malda}).

In the   $D=5$ black hole case  the 
 issue of correspondence between the   supergravity 
and SYM descriptions   
at the first  {\it subleading} order was  previously 
addressed  in  \ci{dps} 
  where the $v^2$ term 
 present in the supergravity potential $\V^{(2)}$ was not 
explicitly determined 
from a  2-loop SYM expression.
We shall demonstrate that  making  a certain natural  choice of
underlying SYM background  and substituting it into the 2-loop 
SYM term $\G^{(2)}$  implicitly determined  by considerations
of  the previous sections
reproduces the {\it complete} expression for the subleading term
 $\V^{(2)}$  in the supergravity potential.

To be able to reproduce the $v^2$ term in 
$\V^{(2)}$ in the case of a 0-brane interacting with 
a bound state reprsenting  the  $D=4$ black hole 
 we shall need 
to  choose a non-abelian ($[F,F]\not=0$) representation 
for the corresponding SYM background and to  include in the 
 the 2-loop SYM  effective action $\G^{(2)}$ 
 additional  `commutator terms' which  were vanishing 
 in all previous examples  described by 
commuting gauge field backgrounds.

\subsection{0-brane interaction with $D=5$  black hole ($0- (4\bot 1\pa 0)$)}

One particular 
 representation for the three-charge $D=5$ black holes with regular horizons
\ci{ts1}  is given by  the $D=11$ configuration of intersecting  \ci{papt}
longitudinal M5-brane and M2-brane with a wave along the common string, i.e. by
$5\bot 2$+wave \ci{ts2}.\foot{There are other possible
configurations, e.g., $2\bot 2\bot 2$ or
 similar $5\bot 2$+wave system with intersection string being
transverse to the 11-th dimension along which there is a finite boost.
 Since  we 
consider the 0-brane charge 
$N_0$  to be  finite  our choice  does not imply any restriction
 (cf. \ci{halyo}).}
Upon dimensional reduction  along the common 11-th dimension
this becomes the $4\bot 1\pa 0$ type IIA configuration
with the 4-brane and the  fundamental string wrapped over  
 a 4-torus  and a circle  which are orthogonal cycles  of the  5-torus. 
 Special 
cases of this  1/8 supersymmetric marginal BPS bound state
 are the 1/4 supersymmetric $1\pa 0$ and $4\pa 0$ states  discussed  in section 4.

As in the previous cases,   to find the supergravity
interaction  potential we 
 shall consider the  action of a 0-brane probe moving in the background
  produced by  this composite source.
The explicit form of the $D=10$ type IIA  
 string-frame    metric,  dilaton and  vector field
representing  the $4\bot 1 \pa 0$  configuration   may  be
 obtained, e.g., by dimensional reduction of the 
$D=11$\  $5\bot 2 +$wave solution  in 
 \ci{ts2}  (cf. \rf{back})
$$
d s^2_{10} =  H_4^{-1/2} H_0^{-1/2} \HH_1^{-1}
 \big[ -dt^2 +   H_0 \HH_1 (dx^2_1+ ... + dx^2_4) 
$$
\be
 +\      H_0 H_4 dx_5^2   
+  H_0  \HH_1 H_4  (dx^2_6 + ...+ dx^2_9)  \big] \ ,     
\la{popw}
\ee
$$e^{2\phi} = \HH_1^{-1} H_0^{3/2} H_4^{-1/2} \ , \ \ \ \ 
A= H_0^{-1}dt \ ,  $$
$$ \HH_1 = 1 + {\Q^{(4)}_1 \ov r^2 }\ , \ \ \ 
H_4=1 +{ Q_4^{(1)}\ov r^2 }\ , 
\ \ \ \ H_0 = {Q^{(5)}_0\ov r^2 }\ , $$
\be 
  \Q^{(4)}_1 = g_s Q^{(4)}_1 =  (2\pi)^{-1/2} g_s L_5 { N_1 \ov N_0}   Q^{(5)}_0
     \ ,     
\la{uyu}
\ee
where the directions $x_1,...,x_4$  are parallel  to the  4-brane and $x_5$  --
to the string.
In  the $D=11$  interpretation  $H_0$  is  the harmonic function of the wave
(we again drop  1 in $H_0$ assuming `null reduction'  \ci{bbpt}).
The  action of 
 a 0-brane     (moving transversely to the internal   5-torus $x_1,...,x_5$)
 is a direct  generalisation of \rf{acci} and \rf{zerr}
 \be
 I_0 = -T_0 \int dt \  H_0^{-1} \bigg( \sqrt {1- H_0 \HH_1 H_4 v^2  }
    -  1 \bigg) \  , 
 \la{herr}
 \ee
 so that the two  leading terms in the 
  interaction potential  are (cf.  \rf{vot},\rf{xxx},\rf{vtwo})
\be
\V^{(1)} =  - { 1 \ov 8 r^{2}} T_0 \bigg[ 4 v^2 ( \Q_1^{(4)} + Q_4^{(1)}) +  
 v^4 Q^{(5)}_0  \bigg]  \ ,  
\la{lead}
\ee
\be
\V^{(2)} =   - { 1 \ov 16 r^{4}} 
T_0  \bigg[ 
  8 v^2 \Q_1^{(4)}  Q_4^{(1)}  +  4 v^4( \Q_1^{(4)} + Q_4^{(1)} ) Q^{(5)}_0
  +   v^6 (Q^{(5)}_0)^2  \bigg] \ . 
\la{vvot}
\ee
To find  a  description of this 
configuration  in terms of a SYM background
 let us note that 
performing T-duality along all of the directions of $T^5$  transforms
$0 \pa 4 \bot 1$ into    
$5\pa 1 +$wave, i.e. a bound state of a D5-brane, D-string and a  wave. A D-string charge on D5-brane may be  represented 
 by an 4d instanton in
$D=5+1$ SYM theory  \ci{doug,vafa} 
 while a momentum wave  --  by 
a plane wave  configuration of SYM fields \ci{dps,malda,gopak}. 

 The explicit representation for  the  momentum wave in $D=6$ SYM theory
is not unique. Since  
 the  SYM  stress tensor contains the two contributions --
of the gauge fields $A_a$ ($a=0,1,...,5$) and of the scalars $X_i$ ($6,7,8,9$), 
  oscillations of both of the  fields may, in general, 
 carry  parts of the total   momentum. This is what we shall assume below,
choosing the following specific background
\be
A_1 = A_1 (\td x_5 +t) \ , \ \ \ \ \ \ \ X_6 = X_6  (\td x_5 +t) \  . 
\la{wve}
\ee
The wave  of $X_6$ 
is  natural to include as it  represents the  momentum in the special case 
of the D-string+wave system, or   the fundamental 
string  winding number 
 in  the T-dual $1\pa 0$ state  considered in section 
4.1.\foot{It is also natural to expect 
 that in  another special case of 
D5-brane+wave state  the momentum should 
 be carried by the transverse ($X_i$) 
oscillations    of the 5-brane.} 

 Adding a 0-brane probe to the  $4\bot 1\pa 0 $ state 
corresponds  in the T-dual picture to  adding  a D5-brane probe 
parallel to the composite  `D5-brane' source. 
As discussed in the previous sections, 
to determine the dependence on  the scalar field background $X_6$ and  
$X_9 \sim vt$ representing 
 the velocity  of the probe 
 we may formally  
 perform  further  T-duality transformations 
along the 6-th and 9-th transverse  directions.  
We then finish with a stationary pure gauge field  $D=8$ configuration 
describing  a 7-brane probe with  a constant electric  flux $F_{09}\sim v $, 
and a 7-brane source
with a 3-brane charge 
(represented by an instanton background  in $1234$ subspace) 
and  a YM   wave
($A_1= A_1 (\td x_5 +t), \ A_6 = A_6 (\td x_5 +t)$) 
carrying  momentum  along $\td x_5$.

Let us first ignore the $X_6$ or $A_6$ background.
Then the $0-(4\pa 0\bot 1)$ configuration 
is described by the superposition 
  of the (anti)self-dual \rf{ggg},\rf{joj}  and $A_1$ wave \rf{ttt}
backgrounds  in the SYM theory on the dual 5-torus, i.e.
 by
the following   $su(n_0 + N_0)$  gauge field strength    (cf.  \rf{ttt},\rf{joj},\rf{jj}) 
\be
F_{09} = v J_0 \ , \ \  \ \ 
F_{12}=F_{34} = q J_1 \ ,  \ \ \ \ 
F_{51} = F_{01} = h J_0  \ , 
\la{tit}
\ee
where $J_0$ and $J_1$ were defined in  \rf{jjj} and \rf{jj}, and 
$q$ is the same as in \rf{joj} while  $h$ is the same as in  \rf{ttt},\rf{noo}, 
i.e.
\be
q^2 = { Q_4 \ov Q^{(4)}_0  } = { Q^{(1)}_4 \ov Q^{(5)}_0  } \ , 
\ \ \ \ \ 
 < h^2> =  { 1 \ov \td L_5} \int d \td x_5\  h^2    
= { \Q_1 \ov Q^{(1)}_0 }   = { \Q^{(4)}_1 \ov Q^{(5)}_0 
}  \ .  
\la{tuuu}
\ee
Including  the $A_6$ background we  get
instead of \rf{tit}
 \be
F_{09} = v J_0 \ , \  \  \ \ \ \ \  \ 
F_{12}=F_{34} = q J_1 \ ,    \la{tata} \ee
\be 
F_{51} = F_{01} =h J_0  \ , \ \ \ \ \ 
F_{56} = F_{06} =w J_0   \   ,  
\la{tat}
\ee
where   the `vector' and `scalar' 
 wave functions $h=  h(\td x_5 +t)$ and $w= w (\td x_5 +t) $  
satisfy a generalisation  of  the second condition  in \rf{tuuu}
\be
  < h^2 + w^2 > = { 1 \ov \td L_5} \int d \td x_5\  ( h^2  + w^2) 
= { \Q_1 \ov Q^{(1)}_0 } = { \Q^{(4)}_1 \ov Q^{(5)}_0 
}  \ .   
\la{tuuut}
\ee
Thus   $ < h^2 + w^2 > $ is proportional to the  momentum 
of the wave in the $5\pa 1 +$wave configuration.

The `instanton+wave'  $u(N)$ gauge field  background  (\rf{tata},\rf{tat} with $v=0$),   which should be
  representing the
marginal BPS bound state  $5\pa 1 +$wave
configuration invariant
under 1/8 of ${\cal N}=2,D=10$  type IIB supersymmetry,
is indeed   preserving the 1/4 of the ${\cal N}=1,D=10$ 
 supersymmetry  of the SYM theory.
The  standard 
condition  of the vanishing of  the gaugino variation is 
\be
F_{ab}^{rs}  \g^{ab} \ep=0  \ ,   \la{ghg}
\ee
where $r,s$ are the internal  $u(N)$ indices and $
\g_{ab} = \g_{[a} \g_{b]}, \  \g_{(a} \g_{b)}=\eta_{ab}=
\diag(-1,1,...,1)$.
As the diagonal matrices $J_0$ and $J_1$ are not proportional, 
choosing  different  values of $r,s$  and combining the resulting equations
we get  two separate
(`instanton' and `wave')  conditions:
\be
 (\g_{12} + \g_{34}) \ep =0\ , \ \ \ {\rm i.e.} \ \ \ 
P_{1234}  \ep  =\ep \ , \ \ \ \ P_{mnkl}  \equiv \ha ( 1 + \g_{mnkl})  \ , 
\la{isti}
\ee
and $ (h \g_1 + w \g_6)(\g_0 + \g_5)\ep=0 $, or 
\be
 (\g_0 + \g_5)\ep=0  \ , \ \ \ {\rm i.e.} \ \ \
P_{05} \ep  =\ep \ , \ \ \ P_{05}  \equiv \ha ( 1 - \g_{05})  \ .  
\la{istik}
\ee
Since the projectors $P_{1234}$ and $P_{05}$ commute , 
we conclude that  the amount of unbroken
supersymmetry is reduced  to  1/4 of the original one.

Computing the  classical  BI  Lagrangian for this   commuting 
 $u(N)$ background  \rf{tata},\rf{tat} one finds 
\be
L = \tr \sqrt { - \det ( \eta_{ab} I + \hat F_{ab} ) } 
= N (1+q^2) \sqrt { 1 -  { h^2  + (1 + q^2)(1+w^2)  \ov 1 + q^2 } \
v^2}  \ .  
\la{yuy}
\ee
This 
provides a test of the marginal  BPS property of the $v=0$  
background: higher-order terms vanish for $v=0$
and the coefficient of the $v^2$ term 
   $\sim h^2 + (1+q^2) (1+w^2)$ has 
its leading-order part being  proportional 
 to the mass ($ M\sim \Q^{(4)}_1 + Q_4^{(1)} + Q^{(5)}_0 
\sim 1+  q^2 + < h^2 + w^2>$) 
of the $5\pa 1+$wave  bound state (see  also \ci{malda}).
The  higher-order correction term $\sim w^2 q^2 v^2$  will be discussed
 below.

Since the  background  \rf{tata},\rf{tat}
involves  only  two  commuting diagonal $u(N)$ 
 matrices, the  
computation of the  1-loop $\hat C_4$  \rf{bbb}
and 2-loop $\hat C_6$ \rf{oonn}   terms in the SYM 
effective action  is essentially  the same as in section 4.
Both $\hat C_4$ and $\hat C_6$ vanish for $v=0$ as expected
for  a  supersymmetric  configuration.
 $\hat C_4$  has the form (cf. 
\rf{resu},\rf{onee})
$$
\hat C_4  = \STr \ C_4 (F) = - {  1 \ov 8} \STr \bigg[ 4 v^2 q^2 J^2_0 J^2_1
+ 4 v^2 ( h^2  + w^2) J^4_0 +    v^4 J_0^4 \bigg]  
$$
\be
 =  \ - { 1 \ov 4}  n_0 N_0 \bigg[ 4 v^2 q^2  + 4 v^2( h^2 +w^2)  + v^4\bigg] \ .
\la{oee}
\ee
Using  \rf{tuuu},\rf{tuuut} we find 
the  agreement  between  the supergravity   \rf{lead}
and the SYM  \rf{eff},\rf{hhh} 
expressions for the leading-order term in the potential.
Note that  the leading-order potential  $\sim \hat C_4$ depends 
 only on  the sum $<h^2 +w^2>$
of the gauge field  and scalar field    contributions to the momentum, 
i.e. it  is not sensitive to how the momentum is distributed between 
the two  terms.\foot{The agreement between the leading-order terms in  $\V$  and $\G$ was 
previously checked  in \ci{dps,malda} where  the 
scalar wave contribution was not included, i.e. $w$ was equal to zero.}
This  will  no longer  be so   for $\hat C_6$. 

For $\hat C_6 = \SSTr C_6 $ we find 
(cf. \rf{resu},\rf{ghgh})
\be
 \hat C_6 = - { 1\ov 16} \SSTr \bigg[  8 v^2  w^2 q^2   J_0^4  J_1^2 
+  2 v^4 q^2  J_0^4  J_1^2  
+    4 v^4 ( h^2  + w^2)  J_0^6    + v^6 J^6_0\bigg]  \ .  
\la{eiie}
\ee
Computing the modified trace $\SSTr$ \rf{rrr},\rf{ttee} as in  \rf{ete}  we get for the 
2-loop coefficient in $\G$ \rf{eff} ($N=n_0 + N_0$) 
$$
 N \hat C_6 = - { 1\ov 8}   n_0 N_0 \bigg[ (n_0 + N_0 )  \big[ 8 v^2  w^2 q^2 +
 2 v^4 q^2  + 4 v^4 ( h^2 +w^2)    + v^6 \big]
$$ \be 
 + \  
N_0 (8 v^2  w^2 q^2   +  2 v^4 q^2)  \bigg] 
   \ . 
\la{ette}
\ee
The   second  $N_0$-term  in the brackets in  
\rf{ette} originates from the 
correction term in $\SSTr$ in \rf{rrr}.
 This  effectively doubles  the coefficients 
of the $J_0^4  J_1^2 $ terms in \rf{eiie}.
The $n_0N_0^2$ part of  \rf{ette} which should be compared with the supergravity potential  becomes
\be
 N \hat C_6  =  - { 1\ov 8}   n_0 N^2_0 \bigg[ 16 v^2  w^2 q^2  + 
 4 v^4  ( q^2+  h^2 +w^2)    + v^6 \bigg]  + O(n_0^2) \ . 
\la{eiee}
\ee
Using  \rf{tuuu},\rf{tuuut} we  conclude that the 
  $v^4$ and $v^6$ terms in the 2-loop term in $\G$ \rf{eff}
are indeed in   agreement with  the classical supergravity 
potential   \rf{vvot}.

As for the $v^2$-term in \rf{eiee} which is present {\it  only}  if $w\not=0$, 
we  find  that it reproduces   the $v^2$ term in  \rf{vvot} provided 
the momentum is  distributed {\it equally} 
 between the gauge field and the  scalar field 
oscillations, i.e. if  (cf. \rf{tuuut}) 
\be
< w^2> = <h^2> = \hal < w^2 + h^2> = \hal  { \Q^{(4)}_1 \ov Q^{(5)}_0 }
\ . 
\la{xzxz}
\ee
It would be interesting  to find  an independent   reason
for imposing this condition. 
  

\subsection{ 0-brane interaction with  $D=4$ black hole 
 ($0- (4\bot4\bot 4\pa 0)$)}

The  1/8 supersymmetric  marginal bound state configurations of $D=11$ theory 
corresponding to $D=4$ extremal black holes with regular horizons
\ci{CYT} may be represented as M-brane  intersections $ 2\bot 2\bot 5\bot 5$ or  
$5\bot 5\bot 5$+wave  wrapped over $T^6 \times S^1$ \ci{KT}.
Assuming that the  intersection direction is the 11-th one, 
the second configuration (which is thus a combination of the three longitudinal 5-branes) admits a simple  SYM  
 description 
in terms of  three `overlapping'
instantons on the dual  6-torus.\foot{There are other  possible SYM (or matrix theory)  embeddings 
of $D=4$ black holes involving finite boosts which represent 
extra parameters of the corresponding non-marginal 
generalisations of the marginal 1/8 BPS bound states  (we need  
not discuss them  here  since  we keep  0-brane number $N_0$ finite
as in \ci{bbpt}). 
As an example, one may consider $2\bot 2\bot 5\bot 5$  with the brane directions being $\{5,6\}, \{4,7\}, \{1,2,3,4,5\}, \{1,2,3,6,7\}$
and add a boost along  direction 1. Reducing  down to $D=10$ along  the boost
direction,  we get  a  non-marginal bound state $0 + (4\bot 4\bot 2\bot 2)$ parametrised
by 5 charges, 
  or, after T-duality along all of 
 the directions of  6-torus, 
$6 + (2\bot 2\bot 4\bot 4)$ (this configuration is  
related to the 5-charge  $D=4$ black hole in \ci{CYT}).}

Dimensional reduction of the $D=11$ background 
$5\bot 5\bot 5$+wave to $D=10$ 
gives the $4\bot 4\bot 4 \pa 0$  configuration  which becomes
$ 2\bot 2 \bot 2\pa 6$ after T-duality along 
the directions of 6-torus.\foot{Other U-dual  $D=10$ configurations like 
 $3\bot 3 \bot 3 \bot 3$  \ci{KT,bal,gree} 
   do  not  have  as simple 
SYM description as the one  existing for 
 $ 2\bot 2 \bot 2\pa 6$  we describe below.
The    $ 2\bot 2 \bot 2\pa 6$ configuration was also 
mentioned  as a  possible
 matrix theory  representation  for $D=4$  black holes in \ci{limart}.}
As we shall demonstrate  below, it
  can be described   in terms of   certain 
1/4 supersymmetric gauge field background in the $D=6+1$ SYM  theory 
on $\td T^6$  which 
may be interpreted as  a  superposition of the 
 three `overlapping' instantons 
(with  the instanton numbers being the 
charges of the three  orthogonal 2-branes). 
The choice of such gauge field  background is not unique, and we shall present 
both `commuting' (see also Appendix) 
and `non-commuting' representations  for it. 

In contrast to the $D=5$ black hole case  considered above
 which does not admit a purely 
D-brane description (one of the charge is
 always from the NS-NS sector),  in the  $ 4\bot 4\bot 4\pa 0$ case 
  all the charges are  from the 
 R-R sector and  so  the correspondence between the  supergravity and SYM 
expressions for the interaction potential should, in principle, be 
 more straightforward to establish. This is indeed so for   the leading-order
term in the  $1/r$ expansion, irrespective of  a particular   choice
of  the gauge field theory  
representation for  $4\bot 4\bot 4 \pa 0$. 
We shall find, however,  that  to be able to  reproduce the subleading term
 in the 
supergravity  potential  (in particular, its $v^2$ part)  one 
 must use the non-commuting version 
of the corresponding  SYM background.

\subsubsection{Supergravity background and  interaction potential}
The $5\bot 5\bot 5+$wave configuration (with 5-brane coordinates $\{1,2,3,4,11\},
\{1,2,5,6,11\}$,  and $\{3,4,5,6,11\}$) 
 reduced  along $x_{11}$ gives the 
 type IIA $4\bot 4\bot 4 \pa 0$
background   with the following 
 string-frame metric, dilaton and 
 R-R  vector field \ci{KT} (cf. \rf{popw}) 
$$
ds^2_{10} = (H_0\Ho \Ht \Htr)^{-1/2}\bigg[-  dt^2 +
H_0 \Htr (dx_1^2+dx^2_2)
+  H_0 \Ht (dx_3^2+dx_4^2) $$ \be  + \
 H_0 \Ho (dx_5^2+dx_6^2)
+  H_0\Ho\Ht\Htr (dx_7^2+dx_8^2+dx_9^2)\bigg] \,,   
\label{fDmetric}
\ee
\be
e^{2\phi}= (\Ho\Ht\Htr )^{-1/2}H_0^{3/2}\,, ~~~~~ \ \ \ \  A=H_0^{-1}dt  \,.
\label{fDdilaton}
\ee
\be
H_0= \frac{Q_0^{(6)}}{r}\,, \ \ \ \ \ ~~~~~H_{4(k)}=1+\frac{Q_{4(k)}^{(2)}}{r} \, ,  \ \  \ \ \ k={ 1,2,3} \ . 
\ee
Here the charges are proportional to the numbers of branes
according to \rf{fif}, i.e.  $Q_0^{(6)} \sim N_0,\  Q^{(2)}_{4(k)}\sim N_{4(k)}$.
The action of a 0-brane probe in  this  background is  again of the same form as 
\rf{pou},\rf{zerr},\rf{herr}\foot{Essentially the same action is found  
for the 
T-dual configuration of  a 3-brane  probe in the 
$3\bot 3\bot 3\bot 3$
background \ci{GKT}.}
\be
I_0 =  -T_0\int dt \ H_0^{-1}\bigg( \sqrt{1-H_0\Ho\Ht\Htr v^2}-1\bigg)
 \ ,  
\label{probact}
\ee
 so that the two  leading terms in the interaction potential \rf{vvv}
are  (cf. \rf{xxx},\rf{vtwo} and \rf{lead},\rf{vvot})
$$
\V^{(1)} =  - { 1 \ov 8 r} T_0 \bigg[   4 v^2 \bigg(Q_{4({\bf 1})}^{(2)}  + 
Q_{4({\bf 2})}^{(2)}  +  Q_{4({\bf 3})}^{(2)}   \bigg) +   v^4 Q^{(6)}_0  \bigg]  
$$ \be
 =-\frac{\pi n_0  }{8r} \bigg[ 4  v^2 \bigg(
\frac{\No}{\Vo}+\frac{\Nt}{\Vt}+\frac{\Ntr}{\Vtr}\bigg)
 + v^6 N_0 \frac{(2\pi)^2}{V_6} \bigg]   \,, \  
\label{Vlead}
\ee
$$
\V^{(2)} =   - { 1 \ov 16 r^{2}} 
T_0  \bigg[ 
  8 v^2  \bigg(Q_{4({\bf 1})}^{(2)} Q_{4({\bf 2})}^{(2)}
+ Q_{4({\bf 2})}^{(2)} Q_{4({\bf 3})}^{(2)}
+ Q_{4({\bf 1})}^{(2)} Q_{4({\bf 3})}^{(2)}\bigg)
$$
$$  +\   4 v^4\bigg( Q_{4({\bf 1})}^{(2)}  + 
Q_{4({\bf 2})}^{(2)}  +  Q_{4({\bf 3})}^{(2)}   \bigg) Q^{(6)}_0
  +   v^6 (Q^{(6)}_0)^2  \bigg]
$$
$$
= -\frac{n_0g_{s}}{  64 (2\pi )^{1/2}  r^2} 
\  \bigg\{ \ha  (2\pi)^2 v^2 \bigg(
{ \No  \Nt \ov  \Vftr }     + 
{   \Nt  \Ntr \ov  \Vfo }    + 
{  \No  \Ntr \ov  \Vft  } \bigg) 
$$
\be +\  4 v^4  N_0  {(2\pi)^{4}  \ov  V_6 }  \bigg(
\frac{\No}{\Vo}+\frac{\Nt}{\Vt}+\frac{\Ntr}{\Vtr}\bigg)
 +v^6  N_0^2 { (2\pi)^{6} \ov V^{2}_6} \bigg\} \,  ,  
\la{Vvot}
\ee
where $V_6 = (2\pi)^6 R_1...R_6$ is the volume of $T^6$, \ 
$\Vo =(2\pi )^2R_5R_6$, $\Vt =(2\pi )^2R_3R_4$,
$\Vtr =(2\pi )^2R_1R_2$ and $V_{4(k)} = V_6/V_{2(k)}$.\foot{Note that these  leading terms in the potential
simplify  when  the tori are self-dual, 
i.e. when $V_p = (2\pi)^{p/2}$: all  
factors  of $2\pi$  inside the brackets cancel out
and one is  left with  combinations of integer numbers of branes 
multiplying powers of velocity.}
Our aim will be to reproduce these expressions by substituting appropriate 
SYM background into  the 1-loop and 2-loop terms in the  SYM effective action \rf{eff}.  Note 
 that the remaining part ($ \sim g^2_s \No  \Nt \Ntr $) 
of the $v^2$ term in the potential  in \rf{probact} 
    which is contained in  
$\V^{(3)}$
should  come from  a  3-loop term in the SYM  effective action.

\subsubsection{SYM backgrounds representing  $4\bot 4\bot 4\pa 0$}

To find a description of $4\bot4\bot4\pa0$  wrapped over $T^6$ 
or of  its T-dual configuration 
$6\pa 2\bot 2\bot 2$  wrapped over $\td T^6$  in terms of a 
  gauge field background  $F_{mn}$ ($m,n=1,...,6$) 
in the SYM theory on $\td T^6$  
one needs to satisfy the following conditions:

(1) $F_{mn}$ 
 should preserve 1/4 of ${\cal N}=1, D=10$ supersymmetry, i.e.
there should exist the corresponding $\ep\not=0$ 
 solution of \rf{ghg};

(2) substituted into the (non-abelian, $U(N_0)$) D6-brane action, 
$F_{mn}$ should induce \ci{doug}  only 
the required charges of the three 2-branes:  it should satisfy  
$\tr F=0,  \ \int \tr (F\wedge F\wedge F) =0$ with 
$ \tr ( F\wedge F) \not=0$  such  that 
 $\int C_3 \wedge \tr ( F\wedge F)$  gives the coupling  of 3-form field 
to the
  charges 
$N_{4 ({\bf 1})}$,$N_{4 ({\bf 2})}$ and $N_{4 ({\bf 3})}$
 of 2-branes wrapped over the  three orthogonal cycles of 6-torus;

(3) the classical BI  Lagrangian  \rf{born}\  
$L =  \Str \sqrt { \det ( \delta_{mn} I +  F_{mn} ) } $
should reproduce the mass of the marginal BPS bound state
$6\pa 2\bot 2\bot 2$, i.e. 
 all higher-order terms in $L_6$  should vanish,   
$L =  \tr ( I + \four F_{mn} F_{mn} )$.

\noindent
The 2-brane charges on a collection of $N_0$  6-branes
  may be represented by   a 
4d (anti)self-dual $SU(N_0)$ 
gauge field  backgrounds. 
There exists several $D=6+1$ SYM backgrounds  which may be interpreted
as  `superpositions' of the three instantons and which 
 satisfy the above conditions and 
 thus  are  candidates for a description  of $6\pa 2\bot 2\bot 2$.

While  all  of them, when substituted into the SYM 
 effective  action \rf{eff}, reproduce the leading term 
 \rf{Vlead} and the $v^4$ and $v^6$ parts of  the subleading term  \rf{Vvot}
in the 
supergravity potential,  it turns out to be impossible to reproduce 
the $v^2$ term in  \rf{Vlead} by using a commuting 
($[F_{mn},  F_{kl}]=0$) background.
Below  and in Appendix we shall  describe  two  
commuting  $F_{mn}$  backgrounds   
which have all the required properties to  represent the 
 $6\pa 2\bot 2\bot 2$  state 
but 
 fail to  give  the $v^2$ term in $\V^{(2)}$.  We shall show that
 there exists a non-commuting 
background which  produces that needed  $v^2$ term  under a natural 
assumption 
that the 2-loop term in the 
SYM effective action \rf{eff}  should, in general, 
contain terms with commutators of $F$.


We shall   consider the following   $su(N_0)$ constant gauge field strength 
 background on $\td T^6$ which may be  viewed  as a generalisation
of the three (anti)self-dual 4d  backgrounds  in  three 
different 4-spaces which intersect over 2-spaces
 \be
F_{14} = - F_{23} = p_1 h_1   \ , \ \ \ \
F_{45} = -  F_{36} = p_2 h_2 \ , \ \ \   \
F_{15} =   F_{26} = p_3 h_3  \ ,  
\la{bb}
\ee
with all other components being zero.
Here $p_k$ are  constants which we shall  fix as 
\be
 p^2_k =  V_{4(k)} \inv  {N_{4(k)}  \ov N_{0}}
 =(2\pi)^2 \td V_{4(k)} \inv  {N_{4(k)}  \ov N_{0}}   \ , 
 \la{pppe} \ee
 while $h_k$ are   independent $su(N_0)$  matrices.
We shall consider the following two different choices for $h_k$.
The first one will be  
\be
h_k = \m_k \otimes I_{{N_0\ov 4} \times {N_0\ov 4}} \ ,   \la{maa} 
\ee
where $\m_k$ are  diagonal $4\times 4$   matrices from the
Cartan subalgebra of $su(4)$
  (used in \ci{taylor} to describe
a YM background representing a non-supersymmetric $6+0$ configuration)
\be
\m_1 = \diag( 1,1,-1,-1) \ , 
\ \ \m_2 = \diag( 1,-1,-1,1) \ , \ \ \m_3 = \diag(1, -1, 1,-1) \ , 
 \la{mati}
\ee
$$ \tr \m_k=0\ , \ \ \ \ \tr (\m_k\m_l) = 4 \delta_{kl}\ , \ \ \ 
\m_k\m_l = |\ep_{kln}| \m_n \ , \ \ \ \ [\m_k,\m_l]=0 \ .  $$
Our second 
choice will be 
\be 
 h_k = \sigma_k \otimes I_{{N_0\ov 2} \times {N_0\ov 2}}   \ , \ \ \ \ \  \ \ 
 \s_k \s_l = \delta_{kl} + i \ep_{kln} \s_n \   , 
\la{pau}
\ee
where   $\s_k$ are   the    $SU(2)$ Pauli matrices.

These two choices of $h_k$  define  
a {\it commuting} and a 
 {\it non-commuting}  $F_{mn}$ backgrounds \rf{bb}  which will have 
the same basic properties.\foot{The commuting background is obviously a solution of the $D=6$  YM
equations. The same should be true also for the non-commuting case, i.e.
there should exist a  potential $A_m$ which solves the classical 
$SU(2)$ YM equations and has 
\rf{bb},\rf{pau} as its field strength.}  
Since the matrices $h_k$ are linearly independent, 
the condition of preservation of supersymmetry \rf{ghg}
 leads  simply to  
\be
(\g_{14} - \g_{23}) \ep =0 \ , \ \  \ \ 
(\g_{45} - \g_{36}) \ep =0 \  , \ \ \  \ 
(\g_{15} +  \g_{26}) \ep =0 \ . 
 \la{gyy}
\ee
The third condition here is a consequence of the first two, 
so that 
 \rf{gyy}   may be expressed in terms
of the two commuting projectors (cf. \rf{isti}) 
$ P_{1234}  \ep  =0, \  P_{3456} \ep  =0 $. As a  result,  
there  exists a solution for $\ep$  representing  the remaining 
1/4 of the  ${\cal N}=1, D=10$ (corresponding to 1/8 of ${\cal N}=2, D=10$)  supersymmetry.\foot{
Related discussions
of supersymmetry-preserving conditions  appeared in \ci{gree,lei,leil,berg}.
Supersymmetric SYM solutions on 8-torus were discussed in \ci{pap}.} 

 It is  also easy to check that  because of the properties of the matrices 
 $\m_k$  or $\s_k$ 
the background  \rf{bb} 
induces only  the required 2-brane charges on the 6-brane,   with 
the three 2-branes  oriented   along the 
$12, 34$ and $56$ cycles of the 6-torus.

Computing the determinant in  the classical BI action \rf{born}
(defined with the symmetrised trace, i.e. ignoring possible commutator terms in non-commuting case   \ci{dbi}) one finds
$$L =  \Str \sqrt { \det { (\delta_{mn} I  + F_{mn} )}} \equiv  \sum_{n=0}^\infty  \Str \ C_{2n} (F)
$$ \be 
= \Str \sqrt { ( I  + p_1^2 h_1^2 + p_2^2 h_2^2  + p_3^2h^2_3 )^2}
=  N_0 ( 1 + p_1^2 + p_2^2 + p_3^2) = \tr ( I + \four F_{mn} F_{mn})  \  . 
\la{ert}\ee
Note, in particular, that  
\be
  C_{2n} (F) =0 \ , \ \ \  n=2,3,...  \ . 
\la{yes}
\ee
As a result,  the  energy  of the
 gauge field configuration  \rf{bb} is, indeed, 
  equal to the  mass of the 1/8 supersymmetric
 marginal bound state $4\bot 4\bot 4 \pa 0$  or $6\pa 2\bot 2 \bot 2$, \foot{ 
See also  \ci{hashi,CT1,taylor} for a similar discussion of 
 energies of $4\pa 0$ 
 and other gauge field configurations.} 
\be
M =   T_6 \int d^6 \td x\  \Str \sqrt { \det { (\delta_{mn} I + F_{mn} )}}
=
 (2\pi)^{1/2} g_s\inv \bigg( N_0 + 
\frac{\No}{\Vo}+\frac{\Nt}{\Vt}+\frac{\Ntr}{\Vtr}   \bigg) \ .
\la{mass}
\ee
In the commuting   $h_k\sim \m_k$ case 
\rf{ert}  follows from the  fact that the 
squares  of $h_k$ are equal to  the   unit matrix $I=I_{N_0 \times N_0}$.
In the  non-commuting   $h_k\sim \s_k$  case
one is to note that $ \Str \sqrt { \det { (\delta_{mn} I  + F_{mn} )}}$
is computed 
 by first expanding the square root  in powers of $F$
 \rf{see} and then applying $\Str$. Since  under the symmetrised trace  
the factors of $F_{mn}$ in $ C_{2n} (F)$ 
 may be treated as commuting, the vanishing of all $C_{2n}, \ n >1, $
which follows from the structure of the abelian version of \rf{bb}, 
 implies  also  that  $\Str\ C_{2n}(F) =0, \ n=2,3,...$.

\subsubsection{Supergravity - SYM   correspondence}

Let us  now demonstrate that while both  commuting \rf{bb},\rf{maa}
(and  \rf{bbbb},\rf{maaa}) and non-commuting 
\rf{bb},\rf{pau}   SYM  backgrounds  (supplemented by the velocity component
and  substituted into the SYM effective action \rf{eff})
reproduce $\V^{(1)}$ 
 \rf{Vlead} and the $v^4$ and $v^6$ parts of   $\V^{(2)}$ \rf{Vvot},
it is only the non-commuting background \rf{bb},\rf{pau}
that  may  generate 
the  subtle $v^2$ term in  \rf{Vvot} provided  also  that 
the  2-loop coefficient $\hat C_6 (F)$ \rf{eeerrr} contains 
commutator terms  like \rf{sixe}. 

The supergravity potential in \rf{Vlead},\rf{Vvot}
expressed in terms of the  SYM background \rf{bb}  takes the following  form
(cf. \rf{xxx},\rf{vtwo})
\be
\V^{(1)} =-\frac{ n_0  }{16  r} \td V_6
 \bigg[ v^2 
{\rm tr} (F_{mn}F_{mn}) + v^4 N_0\bigg]
 \, ,  
\la{feff}
\ee
\be 
\V^{(2)}= -\frac{n_0g^2_{\rm YM}}{(4\pi )^6r^2}\td V_6 
  \bigg[ \ha v^2  ([F,F]^2)
 +v^4  N_0  {\rm tr} (F_{mn}F_{mn}) +v^6 N_0^2\bigg] \, ,  \ \ \ 
\la{ffeff}
\ee
where $g^2_{\rm YM}= (2\pi)^{-1/2} {g_s\td V_6} $ and 
  $([F,F]^2)$  is a notation for  
\be
([F,F]^2) = 16 N_0^2  (p_1^2  p_2^2  + p_2^2  p_3^2  + p_1^2  p_3^2 ) \ . 
\la{fifi}
\ee
The $su(N_0 + n_0)$  SYM background $F_{ab}=( F_{09}, F_{mn}) $
describing the interaction   between 
  a  0-brane  with   velocity $v$    and  
the $4\bot 4\bot 4 \pa 0$  bound state 
  is   given  by  the $su(N_0)$ field strength 
\rf{bb} embedded into $su(N_0 + n_0)$ as 
$F_{mn} \to  \diag(0_{n_0\times n_0}, F_{mn}$ 
and  the `velocity'   component $F_{09}$
(cf. \rf{joj},\rf{jj} and \rf{tata},\rf{tat}) 
\be
F_{09} = v J_0 \ , \ \ \ \ \  \ [F_{09}, F_{mn}]=0 \ , \ \ \ m,n=1,...,6 \ . 
\la{rer}
\ee
Since the indices of the commuting `velocity' 
 and `instanton' parts of $F_{ab}$ 
 do not overlap (in contrast  to what was 
in the case of $D=5$ black hole background \rf{tata},\rf{tat}), 
 one finds 
that the  1-loop and 2-loop coefficients  in \rf{eff}, \rf{bbb},\rf{oonn}
take the following form (cf. \rf{onee},\rf{ghgh})
\be
\hat C_4(F_{ab}) 
 = \hat C_4(F_{mn})  
 - \four \STr \bigg(  v^2 J^2_0 F_{mn}F_{mn} + v^4 J^4_0 \bigg)  \ , 
 \la{ttyyy}
\ee
\be
\hat C_6(F_{ab}) 
 = \hat C_6(F_{mn})   
  -{\textstyle { 1\ov 16}}  \SSTr\bigg(
- v^2 J^2_0 [F_{mn}F_{nk}F_{kl}F_{lm} - \four (F^2_{mn})^2]  
 + \ha v^4 J^4_0  F_{mn} F_{mn} + v^6 J^6_0 \bigg)  \ .
 \la{tyyy}
\ee
As a result,  the  background \rf{bb} (or \rf{bbbb})  substituted into 
the effective action \rf{eff} exactly 
reproduces \rf{feff} and the $v^4$ and $v^6$
terms  in \rf{ffeff}. This is,  of course, not too  surprising
since the  coefficients of the $v^2$ term in \rf{feff} and the  $v^4$ term 
in \rf{ffeff} are   additive in the constituent instanton (4-brane) 
charges, so that 
the agreement is essentially the consequence of the one  in the 
$4\pa0$  case \rf{onee},\rf{ete}.

At the same time, if we assume that $\hat C_6$  is totally symmetric
in the six $F$-factors as in \rf{eeer},\rf{oonn},  then 
the coefficient of the $v^2$ term in \rf{tyyy} 
vanishes and thus does match the one in $\V^{(2)}$ \rf{ffeff}.
Indeed, this coefficient is proportional to $C_4 (F_{mn})$ \rf{fff}
which vanishes identically for the background \rf{bb} (see \rf{yes}).\foot{The same conclusion is reached  in the case of  another  commuting background 
\rf{bbbb}. Since 
$J_0$ is proportional to the  unit matrix on the  
subspace where $F_{mn}$ is non-vanishing,  $ \SSTr [J^2_0 C_4 (F_{mn}) ]  \sim \STr C_4 (F_{mn})$.
Using that the explicit form of  $F_{mn}$
in this case is 
$
F_{12}= q_1 J_1^{\bf (1)} + q_2 J_1^{\bf (2)} , 
\ \ F_{34} =  q_1 J_1^{\bf (1)} + q_3 J_1^{\bf (3)}  ,   \ \ 
F_{56} =  q_2 J_1^{\bf (2)} + q_3 J_1^{\bf (3)} , $
where $ J_1^{(k)}$ are defined as in \rf{jj} with zeroes
in the complementary blocks, we find that 
$C_4 (F_{mn}) = -16 q_1 q_2 q_3  J_1^{\bf (1)}J_1^{\bf (2)}J_1^{\bf (2)}
( q_1 J_1^{\bf (1)} + q_2 J_1^{\bf (2)} + q_3 J_1^{\bf (3)})$
which  does not contain the  required structure in \rf{fifi}
($\sim q^2_1 q^2_2  +... $) even before one takes the traces.}

The non-trivial composite  structure of the $4\bot4\bot4\pa 0$
 bound state  which is 
responsible for the appearance of the  product of the harmonic functions in 
\rf{probact} and thus for   the fact that the 
 $v^2$ term in \rf{Vvot},\rf{ffeff} is 
proportional to  a combination of   {\it products}  of  charges 
of the  constituent 4-branes, 
should,  in fact, be reflected in 
 a  {\it non-abelian}
 nature of the corresponding background
\rf{bb},\rf{pau}.

On general grounds,  the 2-loop term $\hat C_6$ in \rf{eff} may contain various 
$F^6$ commutator terms \rf{sixe}.
To illustrate that they may, indeed,  produce the required 
$v^2 $ term  in \rf{ffeff} let us 
we shall consider  a particular commutator term 
 with  simple  Lorentz
 index  and internal index contractions (cf. \rf{sixe})
\be
\C_6  = i\b_1 
  \Tr (F_{ab} F_{ab} [F_{cd},F_{ef}] F_{cd} F_{ef} ) \ ,  
\la{sixxe}
\ee
where  $\b_1$ is a universal numerical coefficient.  
Using \rf{rer},\rf{bb},\rf{pau}  and noting that
$\Tr (F_{ab} F_{ab} [F_{cd},F_{ef}] F_{cd} F_{ef} ) \to 
 v^2 \Tr (J^2_0  [F_{mn},F_{kl}]^2 )+ ...$
 it is easy to see that this term\foot{Dots stand for $v$-independent terms
that  should cancel  in an appropriate combination of  $F^6$ commutator terms
(the $v=0$ configuration is a BPS one).} 
(multiplied by $N$ according to \rf{eff}) 
contains indeed  the  same $v^2$ 
contribution  as in  \rf{ffeff}, i.e.  is 
 proportional to 
 $ n_0 N_0^2 v^2(p_1^2  p_2^2  + p_2^2  p_3^2  + p_1^2  p_3^2 )$.\foot{Since, e.g.,  $[\s_1,\s_2]=2i\s_3$, the same relation 
is true for the corresponding components of $F_{mn}$ in \rf{bb}
embedded into $su(n_0 +N_0)$, so that  we conclude (cf. \rf{bb} and \rf{joj},\rf{jj})
 that the 
$v^2 p_1^2  p_2^2 $ term in \rf{sixxe}  has coefficient 
$\Tr (J^2_0 J^2_1)$ as in \rf{onee}  which is equal to $2n_0 N_0$.}

\newsection{Concluding remarks}
The  approach we  have used in this paper -- first  extracting an ansatz 
for the leading IR part $\G^{(2)}\sim \int \hat C_6$ of the 2-loop SYM effective  action from  comparison with supergravity potentials  for    
  0-brane interactions  with simple bound states of branes
and then checking its  consistency against  more complicated examples with 
less supersymmetry  may be extended to  various
other cases.  One may consider  
 non-marginal generalisations of, e.g.,  $4\pa 0$  bound state 
(described by a  combination of the  instanton  and 
constant magnetic  backgrounds \ci{guraln})  as well as 
interactions  between two different  bound states of branes
(e.g., $2+0$ and $4\pa0$ with the former treated as a probe as in \ci{CT1}).
One may also study subleading terms in the  interaction potentials for 
 non-supersymmetric configurations  involving  $6+0$ states \ci{taylor}
 or   near-extremal $D=5$ and $D=4$ black holes.  
 This may  lead to   further non-trivial checks of 
the expression for  $\hat C_6$ we have suggested above  and may, in particular, 
  allow  to fix the form of the  commutator terms in it. 

It would be important, of course, to  compute  
the  relevant $F^6$ 
terms in the SYM effective action directly, 
extending the $D=1+0$ result of \ci{BB,bbpt}
to higher dimensions and more general   SYM backgrounds
 and thus   verifying   our conjectures about the structure
of $\hat C_6$.
This may be feasible using a combination of
 techniques  in  \ci{twol,twl,reut}.

Furthermore, it would be  most interesting to 
perform  a string-theory computation of the subleading  (2-loop) terms 
in the interaction potential, checking  that the $r\to 0$ and $r\to \infty$
limits of the string result continue to agree (for relevant supersymmetric
configurations of branes) beyond the  leading 1-loop level considered  in \ci{bachas,dkps,lifmat}. This would provide an explanation for
the supergravity-SYM correspondence at the subleading level  demonstrated in
  \ci{bbpt} and  in  the present  paper.

Finally, there remain also questions about the  role of 
0-branes (and thus of Matrix theory relation) 
in this     correspondence.  
Does  it hold also for other appropriate configurations of branes
with no 0-brane content (or, in T-dual picture, for configurations
other than a  `Dp-brane+...'  parallel to a `Dp-brane + ...')? 
Related question is 
about the role 
of the  large $N_0$ (0-brane number) limit or the 
`null reduction' ansatz  of dropping 1 in the 0-brane harmonic function
\ci{bbpt}.
If there is indeed a weak-coupling string-theory explanation for 
the supergravity-SYM correspondence at the subleading level, it  may  
presumably  apply also  to  some other  
 (nearly) supersymmetric configurations
of (large number of) branes.


\bigskip
\centerline {\ \bf Acknowledgments}
We are grateful to  C. Bachas, M. Douglas and  R. Metsaev  
for useful discussions.
The work of I.C. was supported in part by CRDF grant 96-RP1-253
and  by NSF  grant PHY-9309888. 
A.A.T.  acknowledges  the support
 of PPARC and  the European
Commission TMR programme grant ERBFMRX-CT96-0045.

\setcounter{section}{0}

\appendix{ }
Here we shall describe another representation  for  the $4\bot4\bot4\pa0$
bound state in terms of a  commuting  SYM background on the dual 6-torus
which is different from  the one given in section  5.2.2.

Each 
of  the three longitudinal  M5-branes  in $5\bot5\bot5+$wave configuration 
may be  described \cite{grt,bss}
 by a SYM instanton
on the dual torus, with the instanton charge
being the wrapping number of the  five-brane.
One way to combine them together is to  split the total number of 0-branes 
$N_0$ in $4\bot4\bot4\pa0$ (equal to the number of 6-branes in the T-dual 
$6\pa 2\bot 2\bot 2$ configuration)  into  the three  parts, 
$N_0= \Noo + \Not + \Notr  , $
 and  to consider the three instantons
embedded, respectively, into the   $su(\Noo)$, $su(\Not)$ and $su(\Notr)$
subalgebras of $su(N_0)$.  The 
 components of the  SYM gauge field
potential   on $\td T^6$  may be chosen   as 
$ A_1 = \diag \bigg(   A_1^{({\bf 1})}(\td x_1,\td x_2,\td x_3,\td x_4), \ 
A_1^{({\bf 2})}(\td x_1,\td x_2,\td x_5,\td x_6), 0_{\Notr\times 
\Notr}\bigg)$, $ 
...\ , \ A_6 = \diag \bigg( 0_{\Noo\times \Noo}, \ A_6^{({\bf 2})}(\td x_1,\td x_2,\td x_5,\td x_6), \ A_6^{({\bf 3})}(\td x_3,\td x_4,\td x_5,\td x_6)
\bigg), $ where $ 
A_m^{(k)}\in su ( N_{0(k)}) $   have (anti)self-dual
field strengths, 
\be
F^{(k)}_{mn} = \pm (F^{(k)}_{mn})^* \ , \ \ \ \ \  \ \ 
\frac{1}{16\pi^2}\int_{\td T^4_{(k)}}d^4 \td  x \ {\rm tr}\  (F^{(k)}_{mn}F^{(k)}_{mn})=
 N_{4(k)}\,.   
\la{ppp}
\ee
 Assuming that $N_{0(k)}$ are even,  $F^{(k)}_{mn}$
  may be  taken   in the same explicit  (anti)selfdual form 
 as in 
  \rf{ggg},\rf{gggg} 
$$
F_{12}^{({\bf 1})}=-F_{34}^{({\bf 1})} = q_1  \ 
\sigma_3
\otimes I_{  {N_{0({\bf 1})} \ov 2} \times {N_{0({\bf 1})} \ov 2}}\,,  \ \  \ \ 
\ 
F_{12}^{({\bf 2})}=-F_{56}^{({\bf 2})} = q_2  \ 
\sigma_3
\otimes I_{  {N_{0({\bf 2})} \ov 2} \times {N_{0({\bf 2})} \ov 2}}\,, 
$$ \be  
F_{34}^{({\bf 3})}=-F_{56}^{({\bf 3})} = q_3  \ 
\sigma_3
\otimes I_{  {N_{0({\bf 3})} \ov 2} \times {N_{0({\bf 3})} \ov 2}}\,,  
\la{uyuy} 
\ee
where (cf. \rf{pppe}) 
\be
q^2_k = (2\pi)^2 \td V_{4(k)} \inv  {N_{4(k)}  \ov N_{0(k)}}  \ .
 \la{tyty}
\ee
The choice of the signs 
 here  is  important for preservation of supersymmetry (see below).

The  non-vanishing components of 
 $F_{mn}=\diag (  F_{mn}^{({\bf 1})} , F_{mn}^{({\bf 2})} , 
F_{mn}^{({\bf 3})} )$ are then (cf. \rf{bb}) 
\be
F_{12} = f_1 \ , \ \ \ \    F_{34} = f_2 \ , \ \ \ \ 
F_{56} = f_3 \ ,  
\la{bbbb}
\ee
where $f_k$ are the following  {\it commuting } block-diagonal  
$su(N_0)$   matrices
$$ 
f_1 =   \diag\bigg(  q_1  \sigma_3
\otimes I_{{\Noo \ov 2} \times {\Noo \ov 2}}\ , \ \ 
 q_2 \sigma_3
\otimes I_{{\Not \ov 2} \times {\Not \ov 2}} \ , \ \ 
   0_{{\Notr} \times {\Notr}} \bigg)\  , 
$$ $$
f_2 =    \diag\bigg(   - q_1  \sigma_3
\otimes I_{{\Noo \ov 2} \times {\Noo \ov 2}} \ , \ \ 
 0_{{\Not } \times {\Not}} \ , \ \ 
 q_3 \sigma_3 \otimes I_{{\Notr \ov 2} \times {\Notr \ov 2}} \bigg)\  ,  
$$ 
\be
f_3 =   \diag\bigg( 0_{{\Noo} \times {\Noo}}\ , \ \ 
      - q_2 \sigma_3
\otimes I_{{\Not \ov 2} \times {\Not \ov 2}} \ , \ \ 
- q_3 \sigma_3 \otimes I_{{\Notr \ov 2} \times {\Notr \ov 2}} \bigg) \ .   
\la{maaa}
\ee
To  see  if  this SYM  background preserves  some amount of supersymmetry,  
we write down the condition \rf{ghg} for each of the three diagonal  blocks
of the $su(N_0)$ matrix $F_{mn}$. As a result, we get three  copies of  
 the `single-instanton' condition \rf{isti}
\be
(\g_{12} - \g_{34}) \ep =0 \ , \ \  \ \ 
(\g_{12} - \g_{56}) \ep =0 \  , \ \ \  \ 
(\g_{34} - \g_{56}) \ep =0 \ , 
 \la{gyyp}
\ee
with the third one being a  consequence of the first two.\foot{The choice of all
conditions  in \rf{uyuy} as self-duality ones would  thus 
 lead to a contradiction
unless $\ep=0$ and thus to the  complete breakdown of supersymmetry.} 
The conditions  \rf{gyyp}   may thus  be expressed in terms
of  two commuting projectors 
\be 
P_{1234}  \ep  =\ep \ , \ \ \ \  P_{1256} \ep  =\ep \ ,  
\la{iti}
\ee
implying that 1/4 of the  ${\cal N}=1, D=10$   supersymmetry  is preserved.  

Since   $\tr F_{mn}=0$ and 
$ F_{12} F_{34}  F_{56} =0$ (as follows from  \rf{bbbb},\rf{maaa})
 the only non-vanishing charges induced by this 
background on the 6-brane world volume 
 are the 2-brane charges 
($\sim \int \tr (F_{mn} F_{kl})$)
in the 12, 34 and  56 cycles of the  6-torus. They are  equal 
to $N_{4 ({\bf 1})}$,$N_{4 ({\bf 2})}$ and $N_{4 ({\bf 3})}$.

 The  energy  of this  
 gauge field configuration \rf{bbbb}   obtained from the classical 
BI action is, indeed,   equal to the  mass of the 1/8 supersymmetric
 marginal bound state $4\bot 4\bot 4 \pa 0$.
Since  all components of $F_{mn}$ here are commuting, 
the symmetrised trace in \rf{born}   is equivalent simply  to the 
trace in the fundamental representation and we find the result
similar to that in \rf{ert} 
$$
M =   T_6 \int d^6 \td x\  \tr \sqrt { \det { (\delta_{mn} I + F_{mn} )}}
= \ T_6  \td  V_6    \bigg[ N_0  + N_{0 ({\bf 1})} q^2_1 + 
N_{0 ({\bf 2})}q^2_2 + N_{0 ({\bf 3})}q^2_3\bigg] 
$$ 
\be
= \ T_6 \int d^6 \td x\  \tr  ( I + \four F_{mn} F_{mn} )
 = (2\pi)^{1/2} g_s\inv \bigg( N_0 + 
\frac{\No}{\Vo}+\frac{\Nt}{\Vt}+\frac{\Ntr}{\Vtr}   \bigg) \ .
\la{masss}
\ee

{\vspace*{\fill}\pagebreak}

\end{document}